\documentclass[prd,twocolumn,aps,preprintnumbers,letterpaper]{revtex4-1}
\usepackage{amsmath}
\usepackage{tipa}
\usepackage{amssymb}
\usepackage{graphicx}
\usepackage{xfrac}
\usepackage{bm}
\usepackage{enumerate}
\usepackage{color}
\usepackage[colorlinks = true,
            linkcolor = blue,
            urlcolor  = blue,
            citecolor = blue,
            anchorcolor = blue]{hyperref}
\usepackage{subfigure}
\usepackage{braket}
\usepackage{rotating}
\usepackage{capt-of}
\usepackage{chngcntr}
\usepackage{float}

\newcommand{\dd}{\mathrm{d}}
\newcommand{\eq}[1]{\begin{equation}#1\end{equation}}
\newcommand{\al}[1]{\begin{align}#1\end{align}}
\newcommand{\nn}{\nonumber}

\def\hri#1#2{\href{http://arxiv.org/abs/#1}{[arXiv:#1[#2]]}}
\def\hre#1#2{\href{http://arxiv.org/abs/#1/#2}{[arXiv:#1/#2]}}

\newcommand{\qp}{\vec{q}_\perp}

\begin{document}

\title{The Role of QCD Monopoles in Jet Quenching}

\author{Adith Ramamurti}
\email[]{adith.ramamurti@stonybrook.edu}
\author{Edward Shuryak}
\email[]{edward.shuryak@stonybrook.edu}
\affiliation{Department of Physics and Astronomy, Stony Brook University,\\ Stony Brook, New York 11794-3800, USA}
\date{\today}

\begin{abstract}
QCD monopoles are magnetically charged quasiparticles whose Bose-Einstein condensation (BEC) at $T<T_c$ creates electric confinement and flux tubes.  The ``magnetic scenario" of QCD proposes that scattering on the non-condensed component of the monopole ensemble at $T>T_c$ is responsible for the unusual kinetic properties of QGP. 
In this paper, we study the contribution of the monopoles to jet quenching phenomenon, using the BDMPS framework and  hydrodynamic backgrounds. In the lowest order for cross sections, we calculate the nuclear modification factor, $R_\text{AA},$ and azimuthal anisotropy, $v_2$, of jets, as well as the dijet asymmetry, $A_j$, and compare those to the available data. We find relatively good agreement with experiment when using realistic hydrodynamic backgrounds. In addition, we find that event-by-event fluctuations are not necessary to reproduce $R_\text{AA}$ and $v_2$ data, but play a role in $A_j$. Since the monopole-induced effects are maximal at $T\approx T_c$, we predict that their role should be significantly larger, relative to quarks and gluons, at lower RHIC energies.

\end{abstract}

\maketitle

\section{Introduction}

The Relativistic Heavy Ion Collider (RHIC) at BNL and the Large Hadron Collider (LHC) at CERN provide an abundance of data on a wide range of hadronic collisions, ranging from proton-proton and proton-nucleus collisions to nucleus-nucleus (heavy-ion) collisions. Central and mid-central heavy-ion collisions at sufficiently high beam energies produced a novel form of matter, the quark-gluon plasma (QGP). The explosion of the matter produced in these collisions was found to follow the predictions of relativistic hydrodynamics. Furthermore, even ``small systems" -- central proton-ions and even high-multiplicity $pp$ collisions -- possibly display collective effects in agreement with hydrodynamics.

Such unusual behavior follows from the unexpected kinetic properties of QGP, such as its extremely small viscosity. Another kinetic parameter, to be studied in this paper, is the mean squared momentum transfer per unit length for high energy partons, denoted $\hat q$. As we will see, this transport coefficient also needs to be enhanced as compared to na\"ive perturbative scattering on quarks and gluons.

Jet energy loss is one of the classic signatures of QGP, proposed by Bjorken \cite{Bjorken:1982tu}. Perturbatively produced high-transverse-momentum partons subsequently traverse the QGP medium created in the heavy-ion collisions. These fast moving partons suffer collisional and radiative energy loss, leading to the phenomenon of ``jet quenching."  The {\it nuclear modification factor}, denoted $R_\text{AA}(p_\perp)$,  describes the difference between the spectrum of hard partons that  traverse a medium and those that do not. Its very strong deviation from one, to about 0.3 or so, was among the most  dramatic RHIC discoveries. At the LHC, which produces about twice larger entropy and particle number than RHIC, the nuclear modification was expected to be further enhanced; this, however, has not happened.

Another important property of the in-medium jet energy loss is the azimuthal anisotropy, $v_2(p_\perp)$, the second Fourier component of the expansion of $R_\text{AA}(p_\perp, \phi)$ in azimuthal angle $\phi$. Since the fireball produced in non-central collisions has an elliptical shape in the transverse plane -- the lengths of medium in the direction of the impact parameter $\vec b$ and the orthogonal direction are different --, $v_2$ of jets characterizes the path-length dependence of the energy loss  \cite{Gyulassy:2000gk}. 

Most early models predicted $v_2$ to be approximately twice smaller than what was observed, even when the overall quenching rate was widely varied, and it was speculated that those models were missing some qualitative phenomenon \cite{Shuryak:2001me}. Liao and Shuryak \cite{Liao:2008dk} proposed a possible solution for this discrepancy: a strong dependence of the jet quenching on the matter temperature, with a near-$T_c$ enhancement. 

The possible existence of magnetic monopoles in electrodynamics fascinated leading physicists -- J.J. Thompson, H. Poincar\'e, and especially P.A.M. Dirac \cite{Dirac60} -- for more than a century, but these objects have yet to be experimentally found. With the advent of non-Abelian gauge theories, classical solitons with magnetic charge have been found, by 't Hooft \cite{NUPHA.B79.276} and Polyakov \cite{Polyakov:1974ek} in 1970s. 

These solutions motivated the ``dual superconductor" model of the confinement, proposed by Nambu \cite{Nambu:1974zg}, 't Hooft \cite{NUPHA.B190.455}, and Mandelstam \cite{Mandelstam:1974pi}. In this scenario, monopoles play the role of Cooper pairs and their Bose-Einstein condensation (BEC) at $T<T_c$ expels electric field from the vacuum into confining flux tubes. The detailed justification of this scenario has been obtained by the lattice gauge theory community, who were able to identified the gauge field monopoles and even follow their correlations and motion \cite{D'Alessandro:2010xg,Bonati:2013bga}. Studies of monopole BEC and their contribution to QCD thermodynamics have been recently performed by the authors \cite{Ramamurti:2017fdn}.
 
General arguments based on the renormalization group flow tell us that, moving from hard (UV) to soft (IR) momenta, one should see growth of electric coupling and, due to the Dirac condition ($g_m g_e = 2\pi n$, for some integer $n$ assumed to be 1), the {\em decrease} of the magnetic. So, in a regime where both couplings are comparable, one should expect a comparable density of electric and magnetic quasiparticles. 

It has been argued that such regime occurs {\em above} the phase transition, at $T=(1-2)T_c$. This {\em magnetic scenario} was further used for explaining unusual properties of QGP by Liao and Shuryak \cite{Liao:2006ry,Liao:2007mj,Liao:2008jg}. In this scenario, the uncondensed magnetic monopoles play a dominant role near the QCD critical temperature, $T_c$, where their density peaks.  Scattering between electric and magnetic quasiparticles dominate the transport cross section \cite{Ratti:2008jz}.

The magnetic monopoles were proposed to have a large impact on partons traversing the medium \cite{Liao:2008dk}.  The pioneering studies of their role in these processes have been carried out by Xu, Liao, and Gyulassy \cite{Xu:2015bbz,Xu:2014tda}.
The present paper follows in their steps: the main difference is that, instead of fitting model parameters, we calculate all scattering effects and their consequences directly from first principles. 

One more important observable is the dijet asymmetry. While in $pp$ collisions, parton scattering leads to (back-to-back) dijets which are well-balanced in their transverse momenta, in the presence of matter this balanced is lost, due to differences in the path length and matter fluctuations. 

The amount of literature on jet quenching is significant, and we do not attempt to summarize it; for recent discussion of these difficulties in modeling jet quenching and summaries of progress, see e.g., Refs. \cite{Betz:2014cza, Xu:2014tda, Noronha-Hostler:2016eow, Xu:2015bbz}.

The purpose of this work is to attempt to complement multiple phenomenological models, including scattering on quarks and gluons, as well as on monopoles \cite{Xu:2015bbz,Xu:2014tda}, by a direct calculation of radiative energy loss \cite{Gyulassy:1993hr} from the same lowest-order cross section. We use the Baier-Dokshitzer-Mueller-Peigne-Schiff (BDMPS) framework \cite{Baier:1996sk, Baier:1998yf}, which ascribe the energy loss to gluon radiation caused by  transverse ``kicks'' from the ``scatterers" in the medium. 

We will study the changes in $R_\text{AA}$ and $v_2$ caused by monopoles, including the realistic temperature-dependent monopole densities and monopole correlations. In addition, we will study the effects due to changes in the initial conditions of the medium and the background medium evolution. From our simulations, we will also calculate the dijet asymmetry, $A_j$. Finally, we will make jet quenching predictions for lower energy collisions to be probed in the upcoming Beam Energy Scan II program at RHIC.

\section{Summary of the Jet Energy Loss Formalism}

To remind the reader of the formalism of Refs. \cite{Baier:1996sk, Baier:1998yf}, we briefly review the relevant results, and give expressions for the transport coefficient and cross sections used in the current work.

\subsection{Derivation of the transport coefficient $\hat{q}$ in the BDMPS framework}

The probability for a fast moving quark to have the transverse momentum $q_\perp$ at position $z$, $f(q_\perp^2,z)$ is given by,
\al{
\lambda(z)& \frac{\partial f(q_\perp^2,z)}{\partial z}\nonumber\\&= - f(q_\perp^2,z) + \int{ \frac{1}{\sigma}\frac{\dd\sigma}{\dd^2 \qp^\prime}(\qp^\prime,z)f((\qp-\qp^\prime)^2,z)\dd^2 \qp^\prime}\,,
\label{eq_1}
}
where $\lambda(z)$ is the mean free path of the jet particle and $\sigma$ is the cross section.
Taking the Fourier transforms of $f$,
\eq{
\tilde{f}(b^2,z) = \int{ \dd^2\qp e^{-i \vec{b}\cdot \qp} f(q_\perp^2,z)}
}
and of the potential $V = \frac{1}{\sigma}\frac{\dd\sigma}{\dd^2 \qp}$,
\eq{
\tilde{V}(b^2,z) = \int{ \dd^2\qp e^{-i \vec{b}\cdot \qp} \frac{1}{\sigma}\frac{\dd\sigma}{\dd^2 \qp}(\qp,z)}\,,
}
we can diagonalize the master evolution equation by taking the Fourier transform of the RHS of Eq. \ref{eq_1},
\eq{
\int{ \dd^2 \qp^\prime \int{ \dd^2 b^\prime e^{i (\qp^\prime - \qp)\cdot \vec{b}^\prime} \tilde{f}(b^2,z) \int{\dd^2 b \tilde{V}(b^2,z)e^{-i \vec{b}\cdot \qp^\prime}}}}\,,
}
to get
\eq{
\lambda(z) \frac{\partial \tilde{f}(b^2,z)}{\partial z} = [1 - \tilde{V}(b^2,z)] \tilde{f}(b^2,z)\,.
}

Taking as a model for the potential,
\eq{
V(q_\perp^2) = \frac{1}{\sigma}\frac{\dd\sigma}{\dd^2 \qp}(\qp) = \frac{\mu^2}{\pi(q_\perp^2+\mu^2(z))^2}\,,
}
we find that, at small $b$, the evolution equation for the jet has the form,
\eq{
\frac{\partial \tilde{f}(b^2,z)}{\partial z} = -\frac{b^2}{4}\hat{q}(z) \tilde{f}(b^2,z)\,,
}
The parameter $\hat{q}(z) \equiv \mu^2(z)/\lambda(z)$ is the main property of the matter -- a kind of kinetic coefficient -- that determines all features of the jet quenching.  To get the explicit form for this, we take the Fourier transform of the potential,
\al{
\tilde{V}(b^2,z) = &\int \dd^2\qp  e^{-i \vec{b}\cdot \qp}\left(\frac{\mu^2(z)}{\pi}\right) \frac{1}{(q_\perp^2+\mu^2(z))^2}\nonumber\\
=& \,\,2 \mu^2(z)  \left(\frac{1}{9} b^3 \, _1F_2\left(2;\frac{5}{2},\frac{5}{2};\frac{b^2 \mu(z)
   ^2}{4}\right) \right.
   \\ &\left.+\frac{\pi  (I_0(b \mu(z) )-\mu(z)  b I_1(b \mu(z) ))}{4 \mu^3(z)}\right)\,. \nonumber
 }
 Expanding around $b=0$,
 \eq{
 \tilde{V}(b^2,z) = \frac{\pi }{2 \mu(z) }-\frac{1}{8} \pi  \mu(z)  b^2+\mathcal{O}\left(b^3\right)\,,
}
Then,
\al{
 \frac{[1 - \tilde{V}(b^2,z)]}{\lambda(z)}& \tilde{f}(b^2,z)\nonumber \\
 &\hspace{-1cm}\approx \frac{\frac{1}{8} \left(\pi  b^2 \mu(z) -\frac{4 \pi }{\mu(z) }+8\right)}{\lambda(z)}\tilde{f}(b^2,z)\nonumber \\ 
 &\hspace{-1cm}= - \frac{b^2}{4}\left(\frac{-\pi  b^2 \mu(z) +\frac{4 \pi }{\mu(z) }-8}{2b^2\lambda(z)}\right)\tilde{f}(b^2,z) \label{eq_div} \\ 
 &\hspace{-1cm}= -\frac{b^2}{4}\hat{q}(z)\tilde{f}(b^2,z)\,,
 \label{eq_dv}
 }
 where the transport coefficient $\hat{q}$,
 \eq{
 \hat{q}(z) \equiv \frac{1}{\lambda(z)}\left[\frac{4}{b^2}(1-\tilde{V}(b^2,z))\right] = \frac{\braket{\Delta p^2_\perp(z)}}{\lambda(z)}\,.
 }
 is defined to be the average squared transverse momentum acquired per unit length. This expression is not convergent at $b \rightarrow 0$ (see the expression in parentheses in Eq. \ref{eq_div}), but in logarithmic approximation,
 \al{
 \hat{q}(z) &\approx \frac{1}{\lambda(z)} \int_0^{1/b^2} \dd^2 \qp \qp^2 V(\qp^2,z) \nonumber \\ & =   \rho(z)\int_0^{1/b^2} \dd^2 \qp \qp^2 \frac{\dd \sigma}{\dd \qp^2}(\qp^2,z)\,,
 }
 where $\rho(z)$ is the density of scatterers. This shows that this formalism is equivalent to the transport cross section method of finding the transverse kick; e.g. for a non-relativistic particle traveling in the $z$ direction through the field of a single scatterer, we have that,
\eq{
\Delta p_\perp = \int_{-\infty}^\infty \frac{b \,\,\dd z}{(b^2+z^2)^{3/2}} = \frac{2}{b}\rightarrow \Delta p^2_\perp = \frac{4}{b^2}\,.
}

\subsection{Scattering on electric and magnetic quasiparticles}

The generic form of $\dd \sigma / \dd q_\perp^2$ in QCD is
\eq{
\frac{\dd \sigma}{\dd q_\perp^2} = \frac{C}{(q_\perp^2+\mu^2)^2}\,,
}
For quarks $C_F = 4/3$ and for gluons $C_A = 3$, so, as it is well known \cite{Gyulassy:1993hr, Baier:1996sk}, we have that
\eq{
\frac{\dd \sigma_{qq}}{\dd q_\perp^2} = \frac{(4/3)^2\pi \alpha_s^2(q_\perp^2)}{ (q_\perp^2+\mu_E^2)^2}\,,
}

\eq{
\frac{\dd \sigma_{qg}}{\dd q_\perp^2} = \frac{4\pi \alpha_s^2(q_\perp^2)}{(q_\perp^2+\mu_E^2)^2}\,,
}
and
\eq{
\frac{\dd \sigma_{gg}}{\dd q_\perp^2} = \frac{9 \pi \alpha_s^2(q^2)}{(q^2+\mu_E^2)^2}\,. 
}
Our task at this point is to add scattering on monopoles. Since, for a parton moving ultra-relativistically, the kick from electric and magnetic fields are similar, one  expect the same form of the cross section $\dd \sigma / \dd q^2$, albeit with different factors in the numerator and denominator
\eq{
\frac{\dd \sigma_{qm}}{\dd q_\perp^2} = \frac{(4/3)\pi F^2(q_\perp^2)}{(q_\perp^2+\mu_M^2)^2}\,,
}

\eq{
\frac{\dd \sigma_{gm}}{\dd q_\perp^2} = \frac{3\pi F^2(q_\perp^2)}{ (q_\perp^2+\mu_M^2)^2}\,,
}
with $F(q_\perp^2)$ the monopole form factor. For point-like monopoles, we have that $F(q_\perp) = 1$; for finite-size, we can use the standard treatment of Rutherford scattering in the Born approximation, from which we know that the form factor $F(q_\perp)= \exp\{-q_\perp^2a^2/6\}$ where $a$ is the radius of the scatterer. We only consider long-range
Abelian part of the monopole field, and  do not take into account a more complicated non-Abelian fields in  the monopole core. 

There are no factors of $\alpha_s$ in the monopole cross sections, due to the Dirac condition, which makes their magnitude larger relative to the electric-scatterer cross sections. Another aspect of the parton-monopole cross sections is that the screening mass in denominator, $\mu_M$, is the magnetic screening mass, which, according to lattice measurements, is in QGP about twice smaller than the electric mass, $\mu_E$ \cite{Borsanyi:2015yka}.

Let us, as an exercise, integrate the relevant expressions,
\begin{widetext}
\al{
\int_0^{1/b^2} \dd q^2 q^2 \frac{1}{(q^2+\mu^2)^2}  =-\frac{1}{b^2 \mu ^2+1}+\log \left(\frac{1}{b^2}+\mu ^2\right)-2 \log (\mu )\,,
\label{eq_qgqhat}
}
\al{
\int_0^{1/b^2} \dd q^2 q^2 \frac{\exp\{-q^2a^2/3\}}{(q^2+\mu^2)^2}   =\frac{1}{3} \left(-e^{\frac{a^2 \mu ^2}{3}} \left(a^2 \mu ^2+3\right)
   \left(\text{Ei}\left(-\frac{1}{3} a^2 \mu ^2\right)-\text{Ei}\left(-\frac{1}{3} a^2
   \left(\mu ^2+\frac{1}{b^2}\right)\right)\right)+\frac{3 \mu ^2 e^{-\frac{a^2}{3
   b^2}}}{\frac{1}{b^2}+\mu ^2}-3\right)\,,
   }
\al{
\int_0^\infty \dd q^2 q^2 \frac{\exp\{-q^2a^2/3\}}{(q^2+\mu^2)^2}      = -\frac{1}{3} e^{\frac{a^2 \mu ^2}{3}} \left(a^2 \mu ^2+3\right)
   \text{Ei}\left(-\frac{1}{3} a^2 \mu ^2\right)-1\,.
   \label{eq_integrated}
}
\end{widetext}

The important thing to note is that the $1/b^2$ cutoff has varying effect on the $\hat{q}$ of the monopoles depending on the size of the monopole: for a larger monopole, the energy of the jet does not affect $\hat{q}_m$ as much as $\hat{q}_{q,g}$, which diverge logarithmically with the energy. The lack of logarithmic divergence means that larger monopoles have far less relative effect on high energy quark and gluon jets. Point-like monopoles, on the other hand, behave just as quark and gluon scatterers across all jet energies.

Including $\alpha_s^2(q^2)$ mitigates the logarithmic divergence for the quarks and gluons at high $q^2$, which increases the role of monopoles when scattering high energy jets. In this work, we will only study point-like monopoles, similar to the treatment of \cite{Xu:2014tda}.

While there remains a spread of values of the electric and magnetic screening masses in lattice literature, the general ballpark of those seems to have stabilized over the years. We will follow Ref. \cite{Borsanyi:2015yka}, who carried out large scale simulations with dynamical quarks with realistic masses.  Their results for the magnetic screening mass is $\mu_M/T = 4.48$, and for the electric screening mass $\mu_E/T = 7.31$. 

Note that these values, coming from modern lattice works,  are significantly larger than the ones used before, and particularly in Ref. \cite{Xu:2014tda}. These values lead to a much smaller $\hat{q}$, especially for electric quasiparticle scatterers. This will certainly result in a smaller impact of  quarks and gluons in comparison to models that use pQCD -- or even older lattice -- values for the electric screening mass, as there is a factor of $\mu_E^4$ in the denominator of the transport cross section. 

\subsection{Parton energy loss in the BDMPS framework}

The BDMPS-like energy loss of a parton as it traverses the medium is given by,
\eq{
-\dd E/\dd z \propto  \hat{q}z\,.
}
Then, for our system, we have,
\al{
-\dd E = \,&z\dd z  \frac{\alpha_s N_c}{12} \hat{q}(z,E)  \nn\\
=  \,&z\dd z  \frac{\alpha_s N_c\pi C_p}{12}  \left( \rho_q(z) \int_0^{q^2_\text{max}}\dd q^2 \frac{(4/3)  \alpha_s^2(q^2)}{ (q^2+\mu_E^2(z))^2} \right.\nn \\
& \left.+ \rho_g(z) \int_0^{q^2_\text{max}}\dd q^2 \frac{3  \alpha_s^2(q^2)}{ (q^2+\mu_E^2(z))^2} \right.  \\ 
& \left.+ \rho_m(z) C_\text{corr} \int_0^{q^2_\text{max}}\dd q^2 \frac{1}{ (q^2+\mu_M^2(z))^2} \right)\,, \nn
}
where $z$ is the coordinate parameterizing the line in the transverse plane along which the parton travels, $C_p$ is the color factor of the jet parton, and $C_\text{corr}$ is a correction factor due to monopole correlations, to be determined in Sec. \ref{sec:mono_corr}.

\section{Densities of electric and magnetic quasiparticles }

  \begin{figure}[th!]
\begin{center}
\includegraphics[width=0.5\textwidth,angle=0]{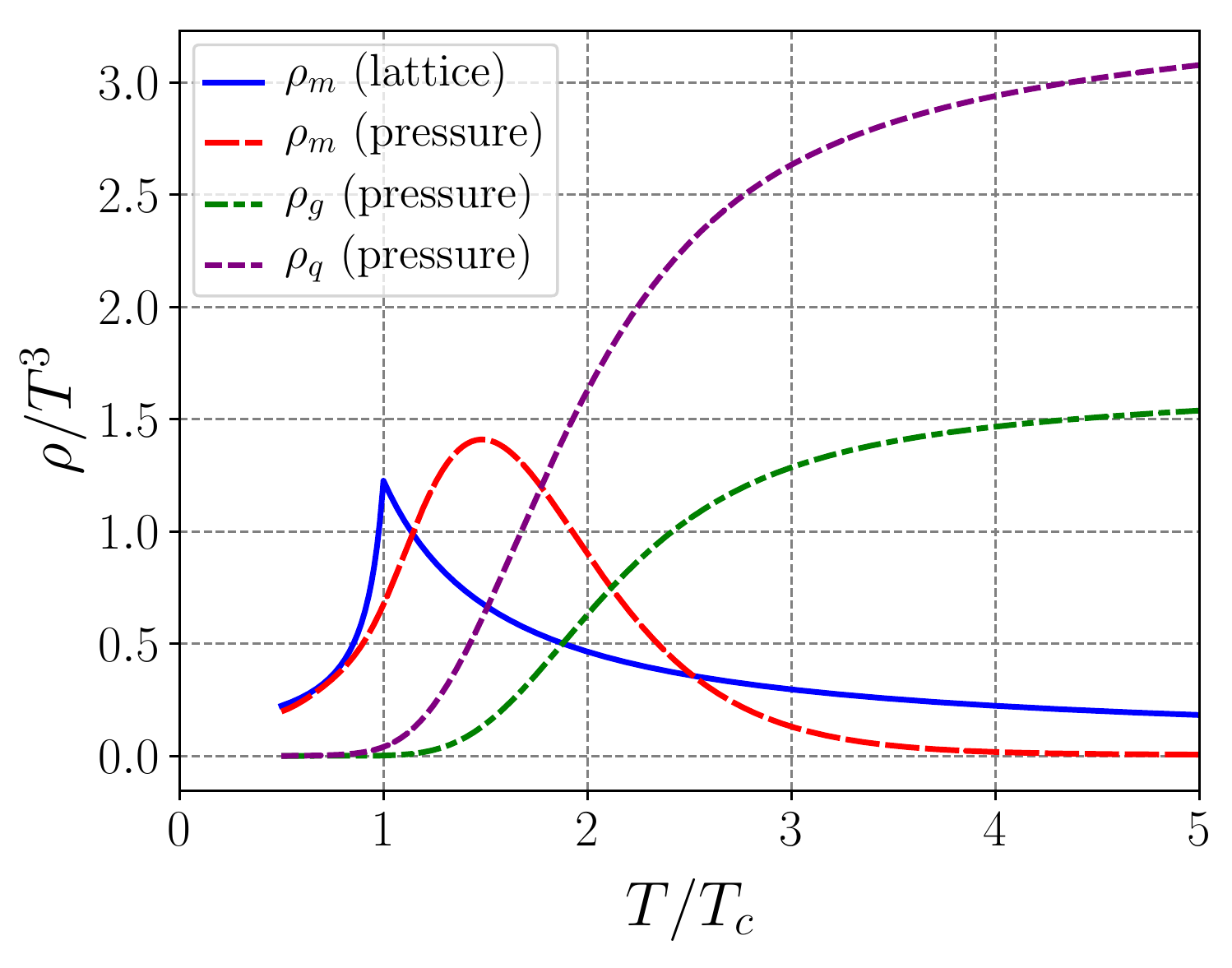}
\caption{ (Color online.) Electric and magnetic quasiparticle densities used. 
The  (blue) solid line shows the magnetic monopole density as directly observed on the lattice. The (red) long dashed line is  the  monopole density extracted from the thermodynamics (pressure), along with the densities of quarks (purple, short dashed) and gluons (green, dot dashed).}
\label{fig:rho}
\end{center}
\end{figure}

For definiteness, in Fig. \ref{fig:rho} we show the densities of gluons, quarks, and monopoles to be used in the calculations below. The details of the analytic fits used are in the Appendix \ref{app_rho}. One should keep in mind that the plotted density is normalized to $T^3$. Such a normalization is appropriate at high $T$, dominated by quarks and gluons, but not necessarily at small $T$. 

In this work, we will use two versions of the monopole density, both obtained from lattice data, but in different ways. The spread of the results is expected to represent the uncertainty existing at the moment. The (blue) solid curve, with a peak at $T_c$, in  Fig. \ref{fig:rho} shows the ``directly observed" monopole density, from Eq. \ref{eq_rho}, which was measured on the lattice \cite{Bonati:2013bga}.

The (red) dashed curve for the density of monopoles, which peaks at about $T\approx 1.5T_c$ rather than at $T_c$, was derived thermodynamically. It is the monopole density needed to reproduce the correct pressure (entropy, energy) of QCD as measured on the lattice \cite{Bazavov:2014pvz}; in the window of temperatures from $1-2T_c$, the energy density, pressure, and entropy density produced by electric quasiparticle degrees of freedom is insufficient. 

We have discussed this thermodynamic estimate in our previous work \cite{Ramamurti:2017fdn}. As we will show below, a monopole density with a peak around $T_c$ seems to be crucial for reproduction of the jet quenching data. 

\section{Correction Due to Correlations of Monopoles} \label{sec:mono_corr}

\begin{figure}[th!]
\begin{center}
\includegraphics[width=.5\textwidth,angle=0]{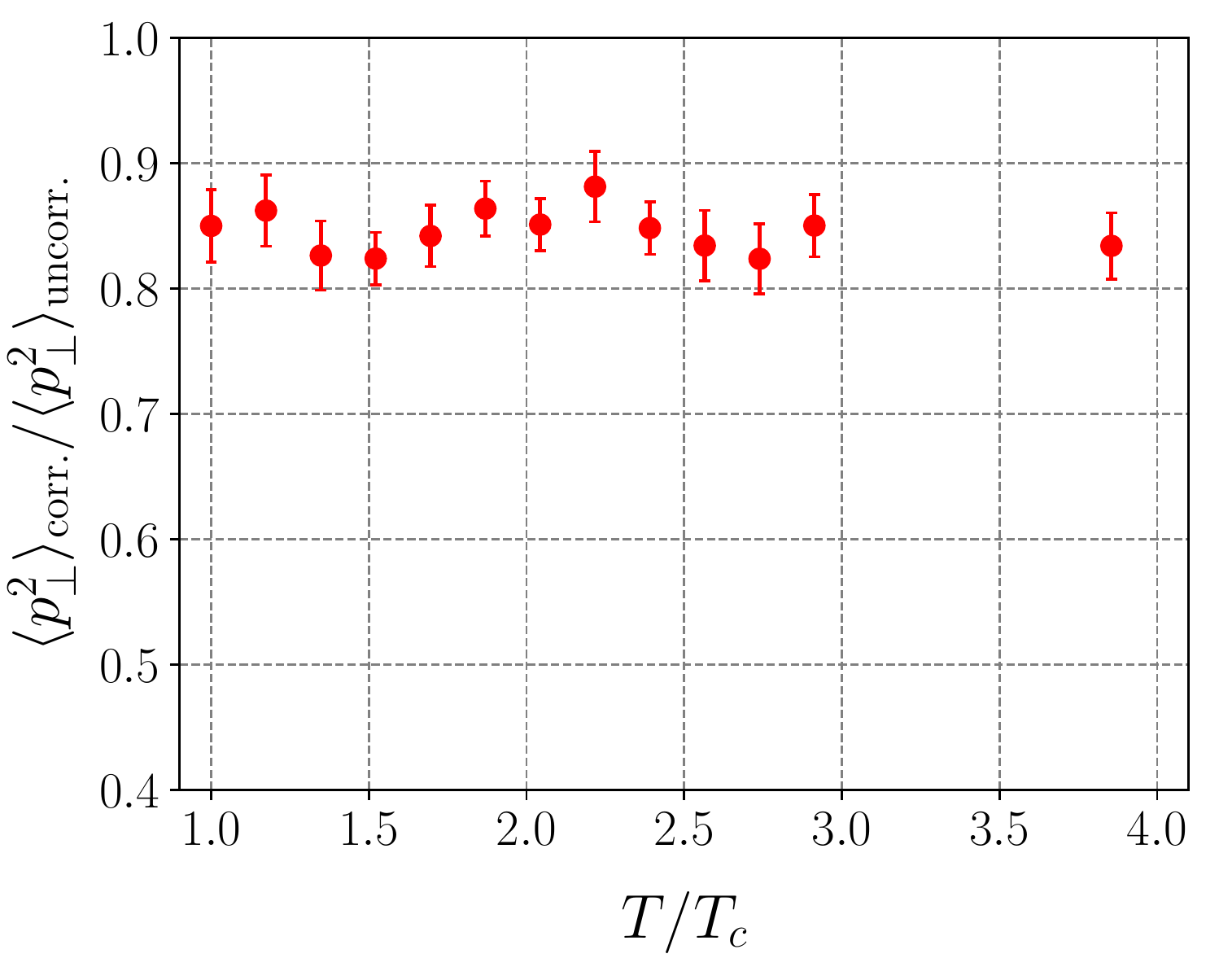}
\caption{ (Color online.) Ratio of correlated to uncorrelated average momentum transfer square per mean free path as a function of the temperature.}
\label{fig:corr}
\end{center}
\end{figure}

Since the magnetic and electric couplings are comparable, the ensemble of magnetic monopoles constitute a strongly coupled plasma in the region of temperatures above $T_c$. In such plasmas, there exist strong correlations between positive and negative charges, which cancel out their fields in some parts of space, reducing their impact on jet quenching.

As expected by the renormalization group flow and Dirac condition, it was directly shown on the lattice (c.f. Refs. \cite{DAlessandro:2007lae, Bonati:2013bga}) that monopoles become more correlated as temperature is increased \cite{Liao:2008jg}. We have evaluated  corrections to the monopole contribution to jet quenching using configurations from our previous path-integral Monte Carlo simulations \cite{Ramamurti:2017fdn}. In that work, we reproduced the lattice correlation functions and the critical condensation of the monopoles, in  a two-component Coulomb Bose gas with varying coupling. In the process of doing these studies, we created quantum ensembles of monopole paths, which we can now use to test what effect these correlations have on the transverse momentum acquired by a jet.

In order to determine the magnitude of this effect, we calculate the net force along a line going through an uncorrelated configuration (random distribution of monopoles and antimonopoles), and then through a random sample of the configurations created in the study of Ref. \cite{Ramamurti:2017fdn}. 

The correlations in the plasma are not extremely strong (there is no crystal like structure, etc.) but are indeed present -- the maximal deviation from 1 of the radial distribution function is 0.2 at 1.1$T_c$ and 0.4 at 3.8$T_c$; see Refs.  \cite{DAlessandro:2007lae, Ramamurti:2017fdn, Bonati:2013bga} for detailed plots of the radial distribution functions.  

\begin{figure}[H]
\begin{center}
\includegraphics[width=.5\textwidth,angle=0]{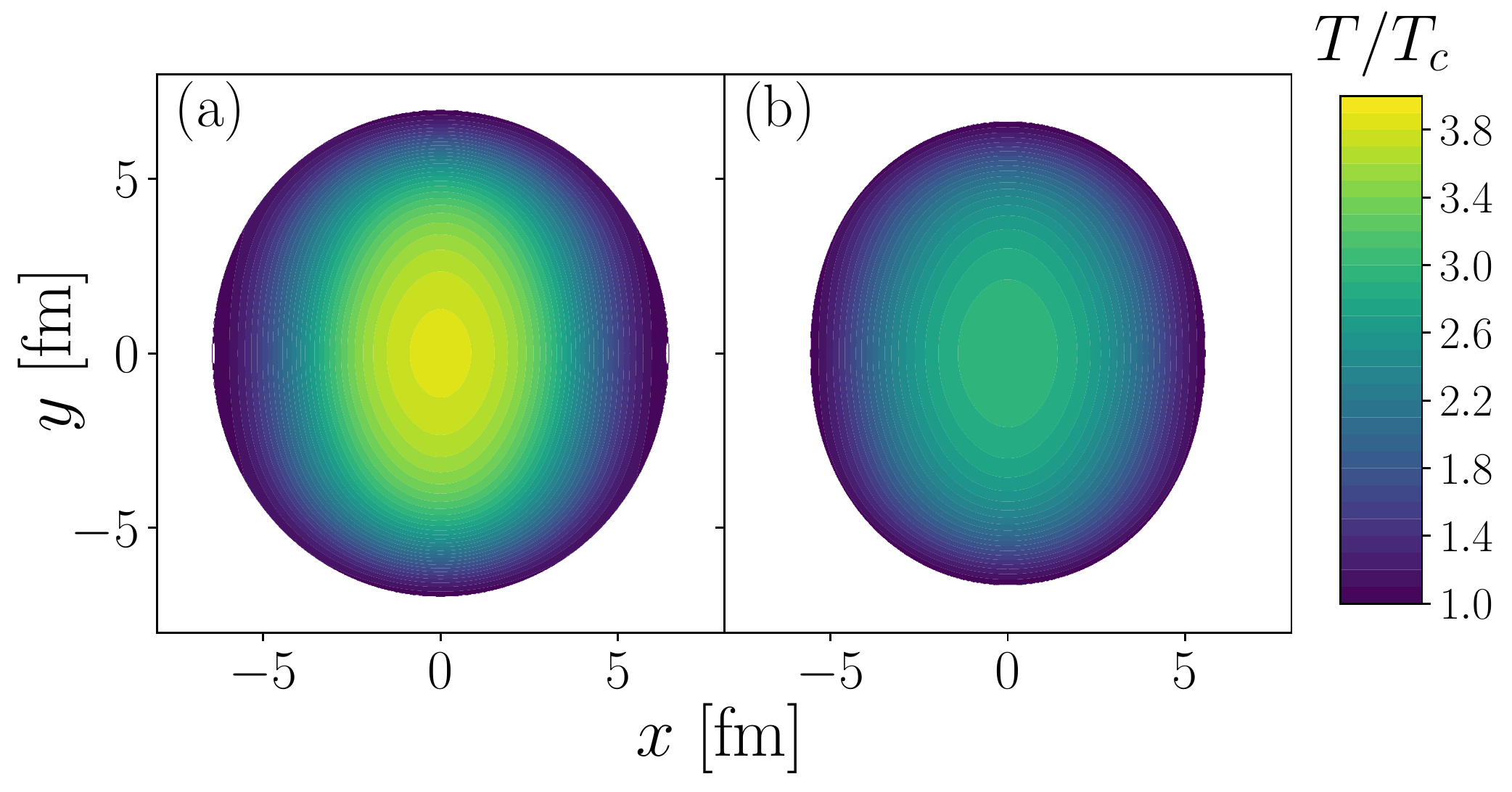}
\caption{ (Color online.) Temperature profile of 20-30\% centrality (a) 2.76 TeV Pb-Pb and (b) 200 GeV Au-Au collisions calculated using the energy density profile at $\tau=0.2$ fm/c from Ref. \cite{Niemi:2015qia} and equation of state from Ref. \cite{Bazavov:2014pvz}.}
\label{fig:tprof}
\end{center}
\end{figure}

\begin{figure*}[th!]
\begin{center}
\includegraphics[width=\textwidth,angle=0]{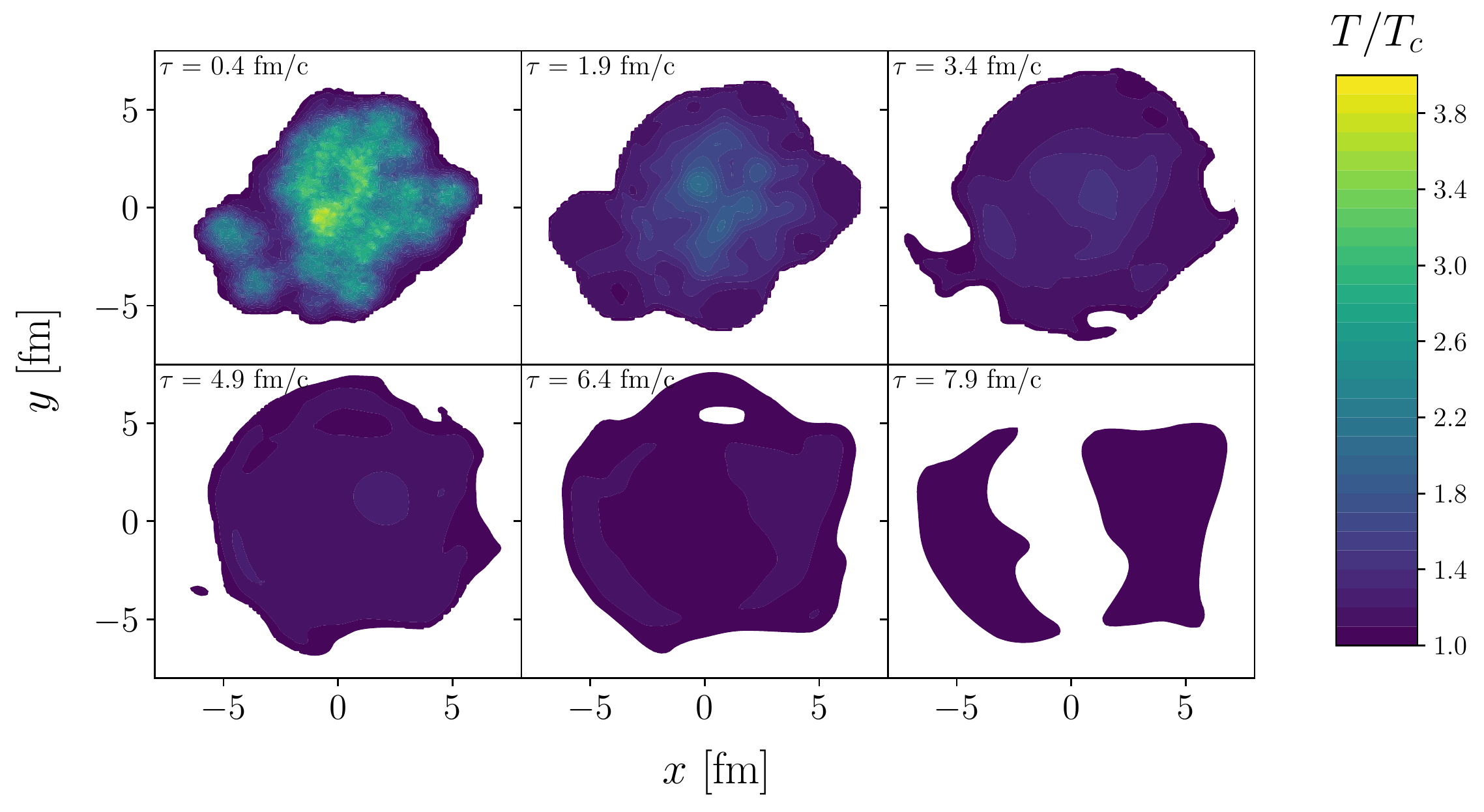}
\caption{ (Color online.) Temperature profile of a 20-30\% centrality 2.76 TeV Pb-Pb collision, shown at various times of the hydrodynamic evolution. This evolution was done using IP-Glasma  initial conditions and (2+1)D hydrodynamics with bulk viscosity.
}
\label{fig:tprofhyd}
\end{center}
\end{figure*}

Fig. \ref{fig:corr} shows the ratio of the average momentum transfer squared per unit length for the correlated and uncorrelated cases. From $T_c$ to 4$T_c$, the ratio is approximately 0.85, meaning that the correlations reduce the $\hat{q}$ by 15\%. Intuitively, the reduction of transferred momentum was expected, since the force on a jet from $+$ and $-$ charges will increasingly cancel the more correlated they are. 

\section{The evolution of the ambient matter at RHIC and LHC energies}

Before we embark on the evaluation of the jet quenching parameters, we need to define  the fireball temperature, energy density, and entropy density profiles. For this study, we will focus on one specific bin, 20-30\%, of centrality, both for LHC and RHIC collisions. Assuming very rapid equilibration, the relation between these profiles are given by equilibrium equation of state (EoS), which has been well studied on the lattice. 

For definiteness, we use parameterization of the energy density from the lattice data of Ref. \cite{Bazavov:2014pvz}, given in Eq. \ref{eq_endens}. The initial energy density distribution corresponding to standard Glauber-type analysis, as in Ref. \cite{Niemi:2015qia}. We also calculated all quantities with IP-Glasma initial conditions, which include fluctuating color fields.

The  temperature profiles of the fireballs at $\tau = 0.2$ fm/c are shown in Fig. \ref{fig:tprof} for both RHIC and LHC energies. One can see that the absolute size and the ellipticity of the near-$T_c$ peripheral regions (blue-purple) are in fact slightly different.

As a first step, we start with simple Bjorken (1+1)D expansion, with the temperature decreasing with time as $T(\tau, x,y) = T(\tau_0,x,y)(\tau/\tau_0)^{-1/3}$.  In a Bjorken-expanding background, the temperature in all regions decrease with time in the same way, and the matter does not expand in the transverse direction.

 We then apply a more realistic (2+1)D hydrodynamic evolution, with both smooth and fluctuating initial conditions. An example of a realistic medium evolution we will use is shown in Fig. \ref{fig:tprofhyd}. One can see that, as time progresses, the (purple) near-$T_c$ region rather quickly takes over the whole fireball, but that the overall size of the fireball region at and above $T_c$ remains approximately the same, unlike what would happen in the (1+1)D Bjorken expansion scenario. Another observation, most clear from two last plots, is that eventually the system splits into two ``nut shells," making the azimuthal asymmetry stronger.

\begin{figure*}[th!]
\begin{center}
\includegraphics[width=\textwidth,angle=0]{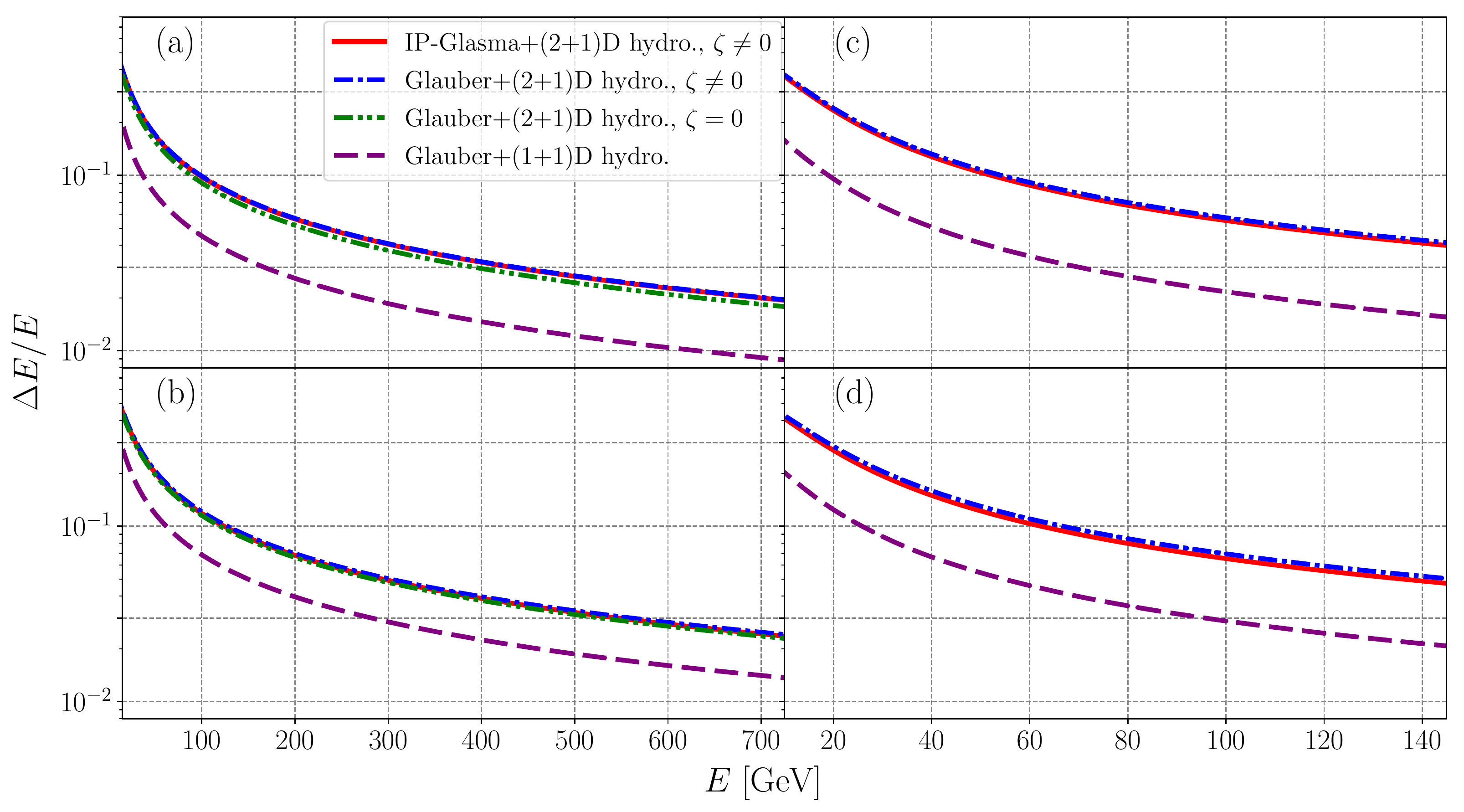}
\caption{(Color online.) Ratio of energy loss to initial parton jet energy in 2.76 Pb-Pb collisions (a),(b), and 200 GeV Au-Au collisions (c),(d). The first row (a),(c) is the results for monopole density from the lattice, while the second row (b),(d) is results for monopole density from the equation of state. The (red) solid curve is for IP-Glasma initial conditions and (2+1)D hydrodynamics with bulk viscosity ($\zeta \neq 0$), the (blue) dash-dot curve is for Glauber initial conditions and (2+1)D hydrodynamics with bulk viscosity ($\zeta \neq 0$), the (green) dash-dot-dot curve is for Glauber initial conditions and (2+1)D hydrodynamics without bulk viscosity ($\zeta = 0$), and the (purple) dashed curve is for the smooth Glauber initial condition with (1+1)D Bjorken evolution.}
\label{fig:dee}
\end{center}
\end{figure*}

As we will see, the hydrodynamic background has an important influence on the results of our jet quenching calculations. As such, it is important that we also reproduce the soft physics of these heavy-ion collisions. The IP-Glasma with hydrodynamics given by MUSIC are studied in Refs. \cite{Ryu:2015vwa,Ryu:2017qzn}, and in general give good agreement with hadronic observables. For our hydrodynamics with optical Glauber initial conditions \cite{Paquet:2017pc}, the simulated and experimental hadronic observables are detailed in Appendix \ref{app_hydro_obs}. 

\section{Jet Quenching at RHIC and  LHC energies: Energy loss, azimuthal anisotropy, and dijet asymmetry}

The probability distributions of quark and gluon jets in their transverse momenta and the location of production were generated by Monte Carlo algorithm in a standard perturbative way, based on Refs. \cite{Spousta:2015fca,Vitev:2005he}. The essential point is that the probability of jet production at a particular location is proportional to the product of two nuclear thickness functions, and that the jet energy spectrum is given by a power law. The produced jets traverse the medium, from the origination point, with an isotropic distribution. 

To calculate hadronic observables from jets, we must apply fragmentation functions to the outgoing quarks and gluons. In this work, we will use the fragmentation functions from Ref. \cite{Kniehl:2000fe} for light quark and gluon jets, going into unidentified charged hadrons and also to neutral pions.

The obtained hadronic spectra are compared to those before traversing the medium, yielding two main observables, the {\em nuclear modification factor},
\eq{
R_\text{AA}(p_\perp, \phi)  =\frac{ \dd N^\text{AA} / \dd p_\perp}{\braket{N_\text{coll}}  \dd N^\text{pp} / \dd p_\perp} \,,
}
and the {\em azimuthal anisotropy}, $v_2$, from
\eq{
\frac{\dd N}{\dd p_\perp \dd \phi} = \frac{1}{2\pi} \frac{\dd N}{\dd p_\perp} \left(1+2 \sum_n v_n \cos(n(\phi-\Psi_n))\right) \,.
}

\begin{figure*}[th!]
\begin{center}
\includegraphics[width=\textwidth,angle=0]{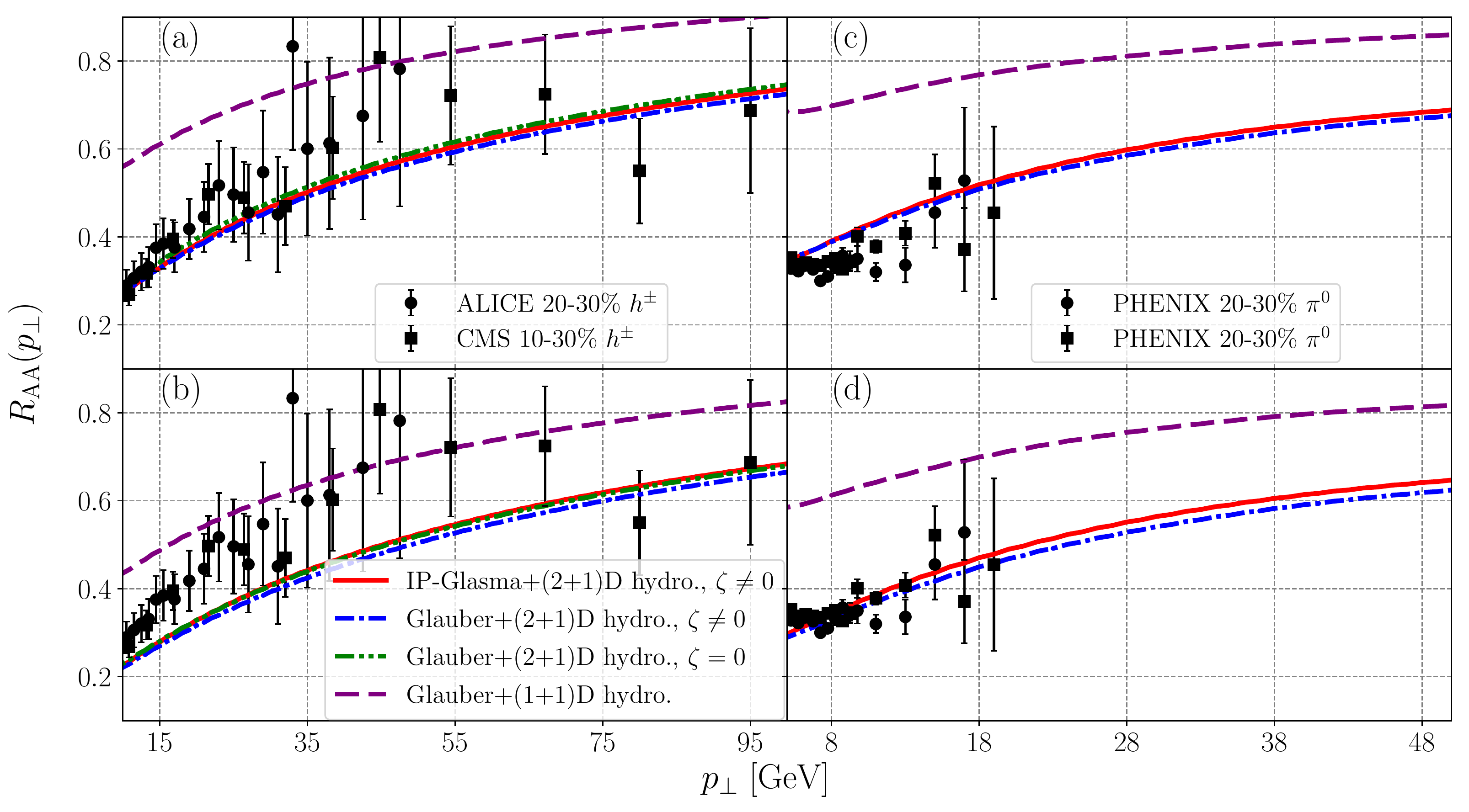}
\caption{ (Color online.) Nuclear modification factor of charged hadrons in 2.76 Pb-Pb collisions (a),(b), and neutral pions in 200 GeV Au-Au collisions (c),(d). The first row (a),(c) is the results for monopole density from the lattice, while the second row (b),(d) is results for monopole density from the equation of state. The (red) solid curve is for IP-Glasma initial conditions and (2+1)D hydrodynamics with bulk viscosity ($\zeta \neq 0$), the (blue) dash-dot curve is for Glauber initial conditions and (2+1)D hydrodynamics with bulk viscosity ($\zeta \neq 0$), the (green) dash-dot-dot curve is for Glauber initial conditions and (2+1)D hydrodynamics without bulk viscosity ($\zeta = 0$), and the (purple) dashed curve is for the smooth Glauber initial condition with (1+1)D Bjorken evolution. Collider data from Refs. \cite{Abelev:2012hxa, CMS:2012aa} for LHC and Refs. \cite{Adare:2008qa,Adare:2012wg} for RHIC.}
\label{fig:raach}
\end{center}
\end{figure*}

\begin{figure*}[th!]
\begin{center}
\includegraphics[width=\textwidth,angle=0]{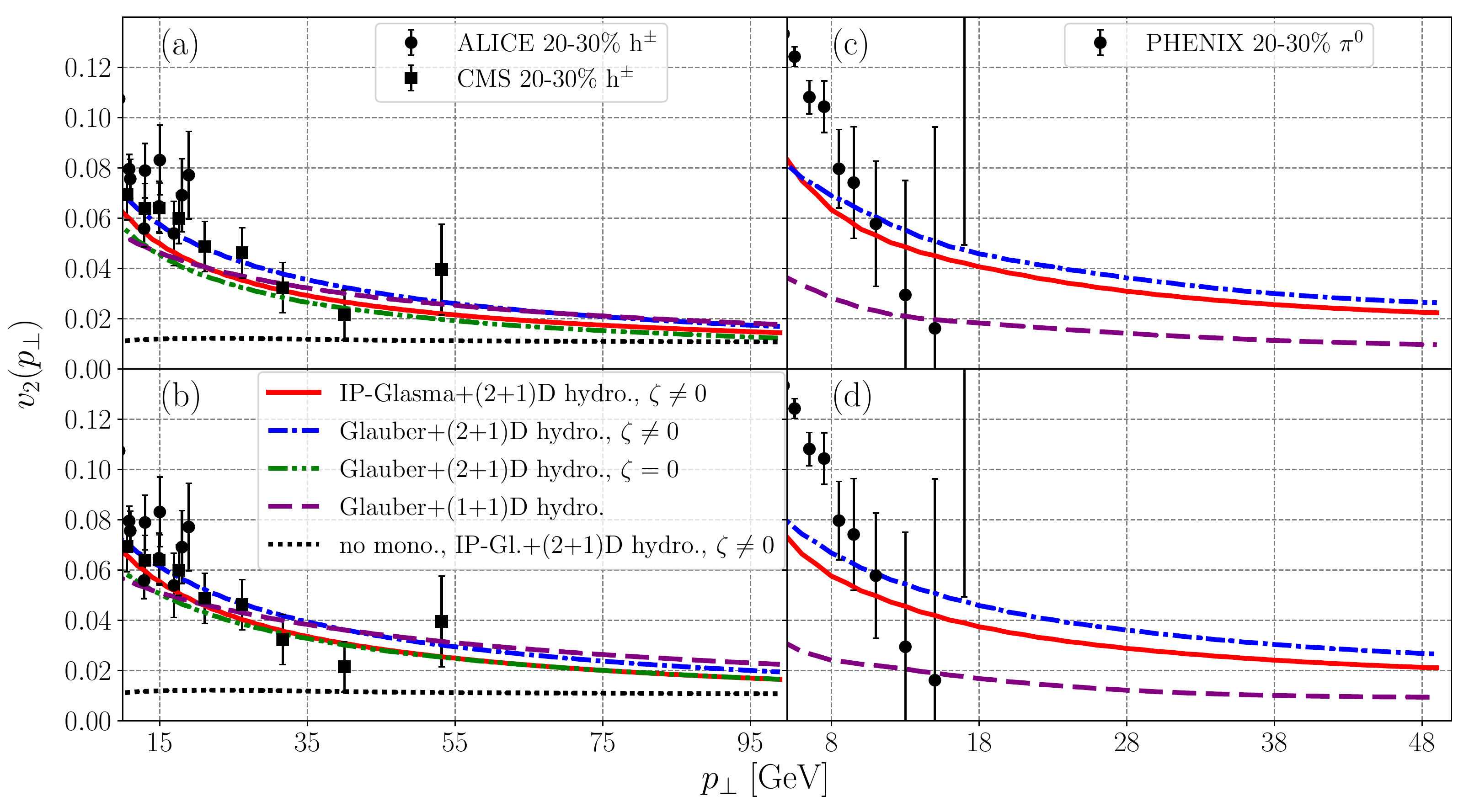}
\caption{(Color online.) Azimuthal anisotropy of charged hadrons in 2.76 Pb-Pb collisions (a),(b), and neutral pions in 200 GeV Au-Au collisions (c),(d). The first row (a),(c) is the results for monopole density from the lattice, while the second row (b),(d) is results for monopole density from the equation of state. The (red) solid curve is for IP-Glasma initial conditions and (2+1)D hydrodynamics with bulk viscosity ($\zeta \neq 0$), the (blue) dash-dot curve is for Glauber initial conditions and (2+1)D hydrodynamics with bulk viscosity ($\zeta \neq 0$), the (green) dash-dot-dot curve is for Glauber initial conditions and (2+1)D hydrodynamics without bulk viscosity ($\zeta = 0$), and the (purple) dashed curve is for the smooth Glauber initial condition with (1+1)D Bjorken evolution. The (black) dotted curve is for for IP-Glasma initial conditions and (2+1)D hydrodynamics with bulk viscosity ($\zeta \neq 0$) with no monopoles. Collider data from Refs. \cite{Abelev:2012di,Chatrchyan:2012xq} for the LHC and Ref. \cite{Adare:2010sp} for RHIC.}
\label{fig:v2ch}
\end{center}
\end{figure*}

The results for $\Delta E / E$ of the jet (prior to fragmentation), $R_\text{AA}$, and $v_2$ of fragmented jets are seen in Figs. \ref{fig:dee}, \ref{fig:raach}, and \ref{fig:v2ch}, respectively. The collider data are from Refs. \cite{Abelev:2012hxa, CMS:2012aa} for LHC 2.76 Pb-Pb TeV $R_\text{AA}$ and Refs. \cite{Adare:2008qa,Adare:2012wg} for RHIC 200 GeV Au-Au $R_\text{AA}$; Refs. \cite{Abelev:2012di,Chatrchyan:2012xq} for the LHC 2.76 TeV Pb-Pb $v_2$ and Ref. \cite{Adare:2010sp} for RHIC 200 GeV Au-Au $v_2$. 

All of the plots are laid out as follows: the first column, comprising subplots (a) and (b), is for LHC 2.76 TeV Pb-Pb collisions, and the second column -- subplots (c) and (d) -- is for RHIC 200 GeV Au-Au. The first row is for the monopole density measured on the lattice, and the second row is for monopole density derived from the equation of state. 

All curves shown are for {\em correlated} monopoles; the effects of correlations on the results are explored in Appendix \ref{app_corr}. The calculations shown are for IP-Glasma initial conditions and (2+1)D hydrodynamics with bulk viscosity, $\zeta \neq 0$ (red, solid curves); optical Glauber initial conditions and (2+1)D hydrodynamics with bulk viscosity, $\zeta \neq 0$ (blue, dash-dot curves); optical Glauber initial conditions and (2+1)D hydrodynamics without bulk viscosity, $\zeta = 0$ (green, dash-dot-dot curves); and optical Glauber initial conditions and (1+1)D Bjorken expansion (purple, dashed curves).

As shown in Fig. \ref{fig:dee}, the Bjorken-evolving background (purple, dashed curve) causes far less energy loss than the scenarios with realistic hydrodynamic backgrounds. This is due to the fact that this is a one-dimensional expansion, and the matter does not expand in the transverse plane; the ellipse of above-$T_c$ medium shrinks inwards to the center of the fireball with time. This is unlike the (2+1)D hydrodynamic case (c.f. Fig. \ref{fig:tprofhyd}), where the size of the above-$T_c$ medium remains approximately constant with time, and all of the medium cools to approximately $T_c$ by $\tau\sim 6$ fm/c. 

\begin{figure*}[th!]
\begin{center}
\includegraphics[width=\textwidth,angle=0]{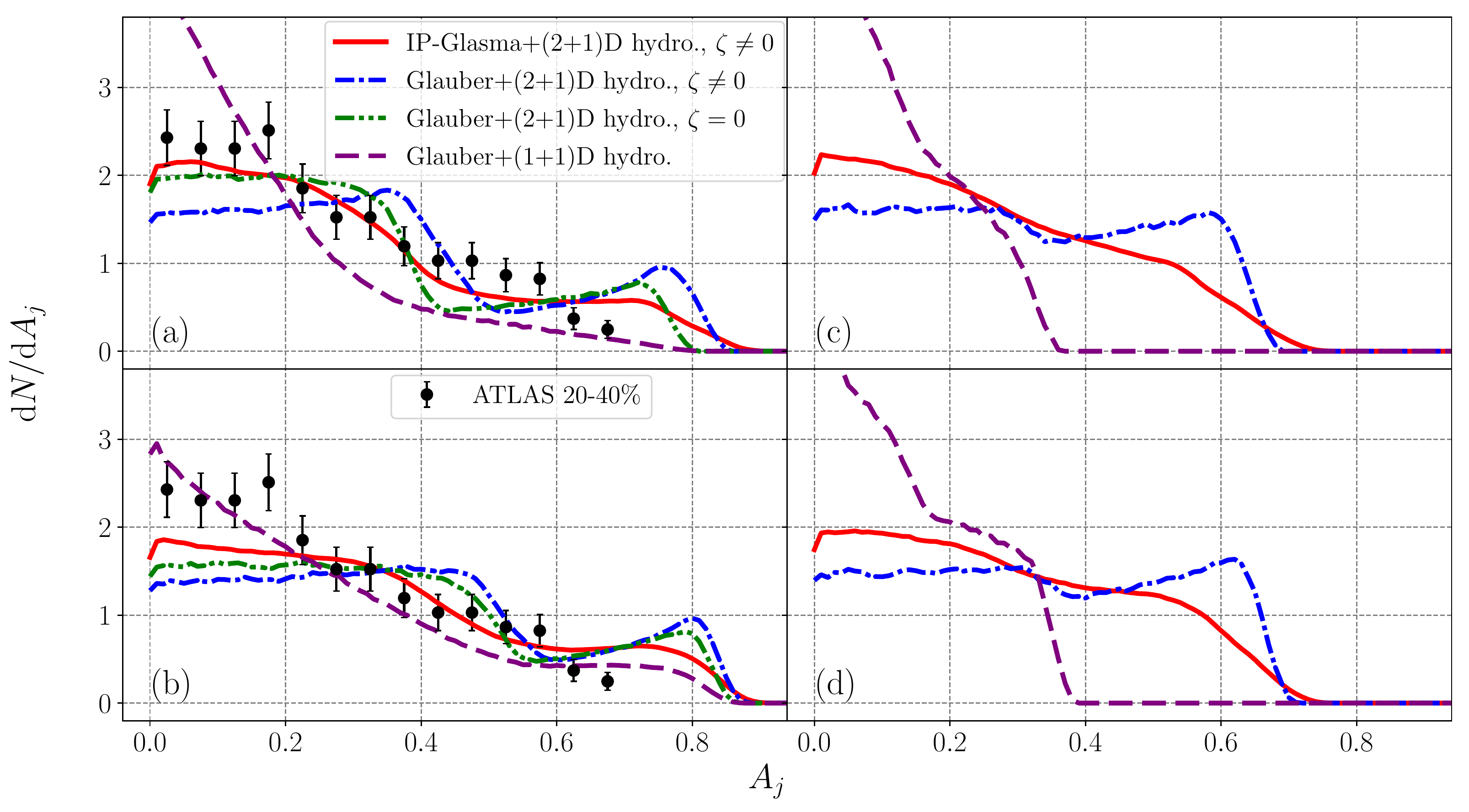}
\caption{(Color online.) Dijet asymmetry of parton jets in 2.76 Pb-Pb collisions (a),(b), and 200 GeV Au-Au collisions (c),(d). The first row (a),(c) is the results for monopole density from the lattice, while the second row (b),(d) is results for monopole density from the equation of state. The (red) solid curve is for IP-Glasma initial conditions and (2+1)D hydrodynamics with bulk viscosity ($\zeta \neq 0$), the (blue) dash-dot curve is for Glauber initial conditions and (2+1)D hydrodynamics with bulk viscosity ($\zeta \neq 0$), the (green) dash-dot-dot curve is for Glauber initial conditions and (2+1)D hydrodynamics without bulk viscosity ($\zeta = 0$), and the (purple) dashed curve is for the smooth Glauber initial condition with  (1+1)D Bjorken evolution. Collider data from Ref. \cite{Aad:2010bu}.}
\label{fig:dijet}
\end{center}
\end{figure*}

Therefore, in the Bjorken-evolving case, the parton jet ``sees'' far less medium during its traversal of the fireball, causing less energy loss. On the other hand, the Glauber and IP-Glasma initial condition models lead to a larger energy loss. This loss is very similar with all initial conditions, provided that the hydrodynamic evolution is realistic.

Since $R_\text{AA}$ is another measure of medium-induced energy loss, the results are very similar to that of $\Delta E/E$. Shown in Fig. \ref{fig:raach}, the Glauber and IP-Glasma initial conditions coupled to realistic hydrodynamic models all agree with each other, while the Bjorken evolution gives a much larger result that is incompatible with the experimental data. The LHC 2.76 TeV Pb-Pb $R_\text{AA}$ data is better fit by the monopole density measured on the lattice (upper left panel), compared to that given by the equation of state (lower left panel). On the other hand, the RHIC 200 GeV Au-Au $R_\text{AA}$ is fit quite well by both monopole densities when using realistic hydrodynamic backgrounds.

The azimuthal anisotropy, is shown in Fig. \ref{fig:v2ch}. Without monopoles -- the black dotted line on the left panels -- the $v_2$ is roughly .015, much smaller than the experimental data.   The $v_2$ results for both monopole densities roughly agree with LHC 2.76 TeV Pb-Pb $v_2$ data when using hydrodynamic backgrounds, including Bjorken evolution.  The $v_2$ for RHIC 200 GeV Au-Au data is not in complete agreement for the (2+1)D hydrodynamic models (Glauber and IP-Glasma initial conditions); the slope is less steep in the model than in the data, although the order of magnitude is correct. On the other hand, the model disagrees strongly with Bjorken expansion.

In all the preceding discussion, we see that, with minor variation, the $\Delta E/E$, $R_\text{AA}$, and $v_2$ obtained with Glauber and IP-Glasma initial conditions and realistic hydrodynamics (blue dash-dot, green dash-dot-dot, and red solid curves) all agree with each other. This leads us to believe that initial-state fluctuations play only a small role in these quantities and that event-by-event analysis is not necessary, which is opposite to what is claimed in Ref. \cite{Noronha-Hostler:2016eow}; the inclusion of monopoles in our jet quenching framework allows for the simultaneous description of $v_2$ and $R_\text{AA}$ even when using smooth initial conditions. 

This most likely occurs because our model is most sensitive to near-$T_c$ medium, which is on the periphery at early times; at later times, when the whole fireball is near-$T_c$, most of the initial state fluctuations are more-or-less damped out. If a model for jet energy loss focuses on early times or is sensitive more to medium at higher temperatures, then the fluctuations would play a much larger role.

In addition, whether or not the hydrodynamic evolution has bulk viscosity does not make a large difference; the variation between the Glauber initial condition with and without bulk viscosity can be explained by the difficulty in tweaking parameters producing the correct hadron yield without bulk viscosity, giving a different background for the jet to traverse (see Appendix \ref{app_hydro_obs}).

This discrepancy in $R_\text{AA}$ between the two forms of monopole density  -- the EoS density not working for 2.76 TeV collisions and both the lattice and EoS densities working for 200 GeV collisions -- can be explained by the difference in high-temperature behavior. While both have relatively similar peaks near $T_c$, the pressure scheme density goes to zero by $T/T_c=4$, and is less than the lattice scheme by $T/T_c=2.5$; see Fig. \ref{fig:rho}. 

On the other hand, the lattice measured density does not go to zero at high temperature, but rather falls off as $\rho/T^3 \sim \log(T/T_c)^{-3}$. As a result, at higher energy collisions (where the temperature is higher at initial time), the variant with directly observed lattice monopoles gives a different monopole contribution than the thermodynamical fit, while in the lower energy collisions, the contributions end up being similar. 

\begin{figure}[th!]
\begin{center}
\includegraphics[width=.5\textwidth,angle=0]{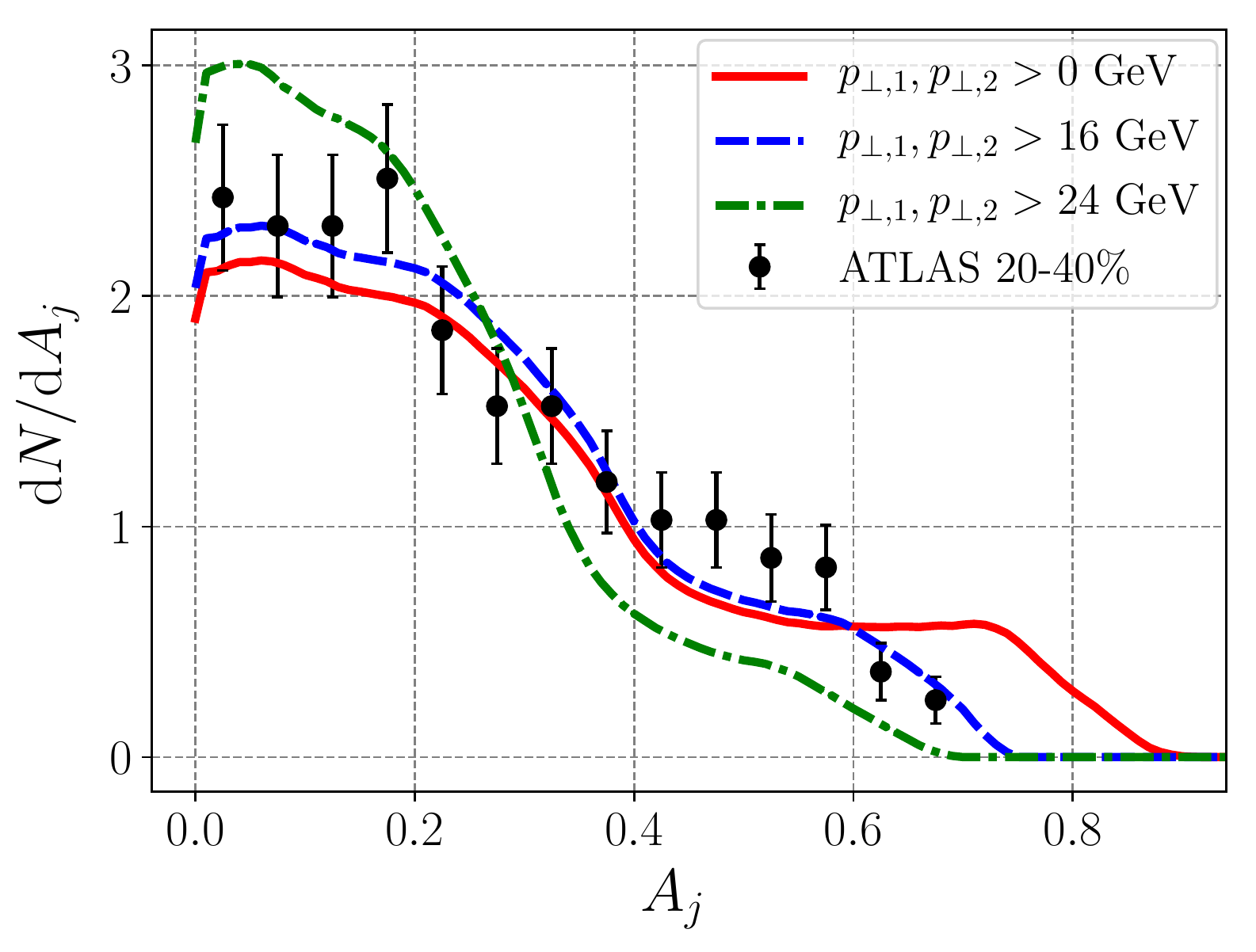}
\caption{(Color online.) Dijet asymmetry of parton jets with no $p_\perp$ cut (red, solid curve), $p_\perp>16$ GeV (blue, dashed curve), and $p_\perp>24$ GeV (green, dot-dashed curve) in 2.76 Pb-Pb collisions for the IP-Glasma initial conditions with (2+1)D hydrodynamics with bulk viscosity and monopole density given by the lattice. Collider data from Ref. \cite{Aad:2010bu}.}
\label{fig:dijetcut}
\end{center}
\end{figure}

At very low energies, where the initial temperature is just around $T_c$, the sensitivity of the jet quenching parameters to the monopole density is the highest. This means, that if the future PHENIX experiment at RHIC will be able to study jet quenching in the low energy range, we will get better understanding of the location of the monopole density peak. Understanding of this quantity is key in determining the effects of monopoles on parton jets. 

In our simulations, we also generated back-to-back partons of the same initial energy, and evaluated the energy lost by each of them. The difference is known as the {\em dijet asymmetry}, and is characterized by,
\eq{
A_j = \frac{|E_1-E_2|}{E_1+E_2}\,.
}
This quantity is plotted in Fig. \ref{fig:dijet}. The data for LHC 2.76 TeV Pb-Pb are from Ref. \cite{Aad:2010bu}. 

We first note that the Glauber and IP-Glasma initial conditions with realistic hydrodynamics (blue dash-dot, green dash-dot-dot, and red solid curves) all roughly follow the experimental data. There is more distinction between the different initial conditions in this quantity, and the best agreement seems to be with the IP-Glasma initial condition and, as before with $R_\text{AA}$, with the monopole density as measured on the lattice. In the case of the dijet asymmetry, we find that initial-state fluctuations play a role in explaining the distribution of energy loss asymmetry between back-to-back parton jets. 

Fig. \ref{fig:dijet} also shows a distinctive bump at $A_j \approx 0.8$ for LHC energies and $A_j \approx 0.6$ for RHIC energies. This bump is due to the fact that we have included jets with low initial energy in our calculations. An asymmetry of $A_j=1$ would indicate that one jet was completely quenched by the medium. The bump comes from the fact that, for low energy back-to-back jets, there is a significant likelihood that the relative difference in final energies is larger, leading to a larger $A_j$. Since low energy jets are much more probabilistic than high energy jets, these events accumulate. 

Therefore, the location of this bump and its size are dependent on the low-energy limit of the produced jets in simulations and any imposed cuts on final jet energies in both simulations and experiment. We also note that the IP-Glasma initial conditions give less pronounced of a bump than other initial conditions.

This bump can, in principle, be (re)moved by a final parton energy $p_\perp$ cut, as is done in experimental analyses. Fig. \ref{fig:dijetcut} shows an example of this for the results of the IP-Glasma+(2+1)D hydrodynamic simulations of 2.76 TeV Pb-Pb collisions and monopole density given by the lattice, for cuts $p_\perp > $ 16 GeV (blue dashed curve) and a $p_\perp > $ 24 GeV (green dash-dot curve). The bump at large $A_j$ disappears and the general shape of the curves are altered; this happens for all curves and at LHC and RHIC energies. 

It is clear that the dijet asymmetry is very sensitive to the range of energies selected for the back-to-back jets. In particular, the $p_\perp > 16$ GeV cut shown in Fig. \ref{fig:dijetcut} (blue dashed curve) makes the calculation with the most realistic background, the IP-Glasma fluctuating initial conditions, roughly follow the experimental data. 

We conclude that our model, with realistic initial conditions, can qualitatively produce the correct behavior of the dijet asymmetry. The response of our data to the adjustment the $p_\perp$ cuts shows the sensitivity of the observable to various parameters introduced in both theoretical and experimental analysis, and therefore requires much more study.

\begin{figure}[th!]
\begin{center}
\includegraphics[width=.5\textwidth,angle=0]{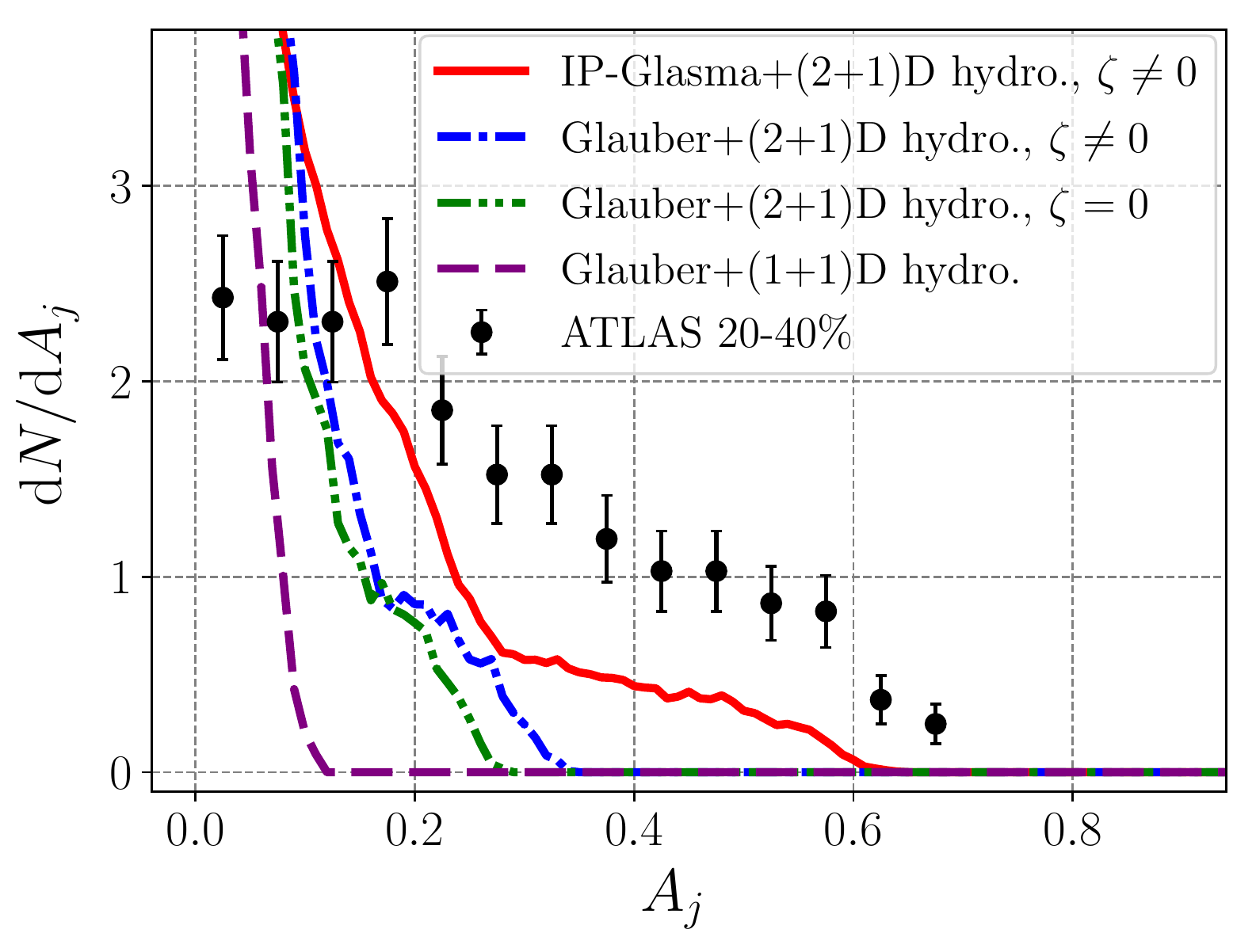}
\caption{(Color online.) Dijet asymmetry of parton jets originating in the center of the fireball in 2.76 Pb-Pb collisions with the monopole density from the lattice. The (red) solid curve is for IP-Glasma initial conditions and (2+1)D hydrodynamics with bulk viscosity ($\zeta \neq 0$), the (blue) dash-dot curve is for Glauber initial conditions and (2+1)D hydrodynamics with bulk viscosity ($\zeta \neq 0$), the (green) dash-dot-dot curve is for Glauber initial conditions and (2+1)D hydrodynamics without bulk viscosity ($\zeta = 0$), and the (purple) dashed curve is for the smooth Glauber initial condition with  (1+1)D Bjorken evolution. Collider data from Ref. \cite{Aad:2010bu}.}
\label{fig:dijet_cent}
\end{center}
\end{figure}

Shown in Fig. \ref{fig:dijet_cent} is the result for the dijet asymmetry $A_j$ selecting only jets that start in the center of the fireball, for 2.76 TeV Pb-Pb collisions and the monopole density found on the lattice. The data is definitively {\em not} reproduced by any of the curves, which shows that the asymmetry, in this model, comes from the path-length difference between the trigger and secondary jets, rather than from the fluctuations in the matter or in the fragmentation processes.

We do not take into account relativistic effects of the fluid flow velocity on the jet. This was studied in Appendix A of Ref. \cite{Xu:2015bbz}; the authors concluded that, for $R_\text{AA}$ and $v_2$, this correction was negligible. 
This is expected since our model only takes into account the instantaneous impact parameter between the jet parton and the scatterer; the cross sections depend on momentum transfer $t$ but not on the energy $s$.

\section{Predictions for the Beam Energy Scan}

\begin{figure}[th!]
\begin{center}
\includegraphics[width=.5\textwidth,angle=0]{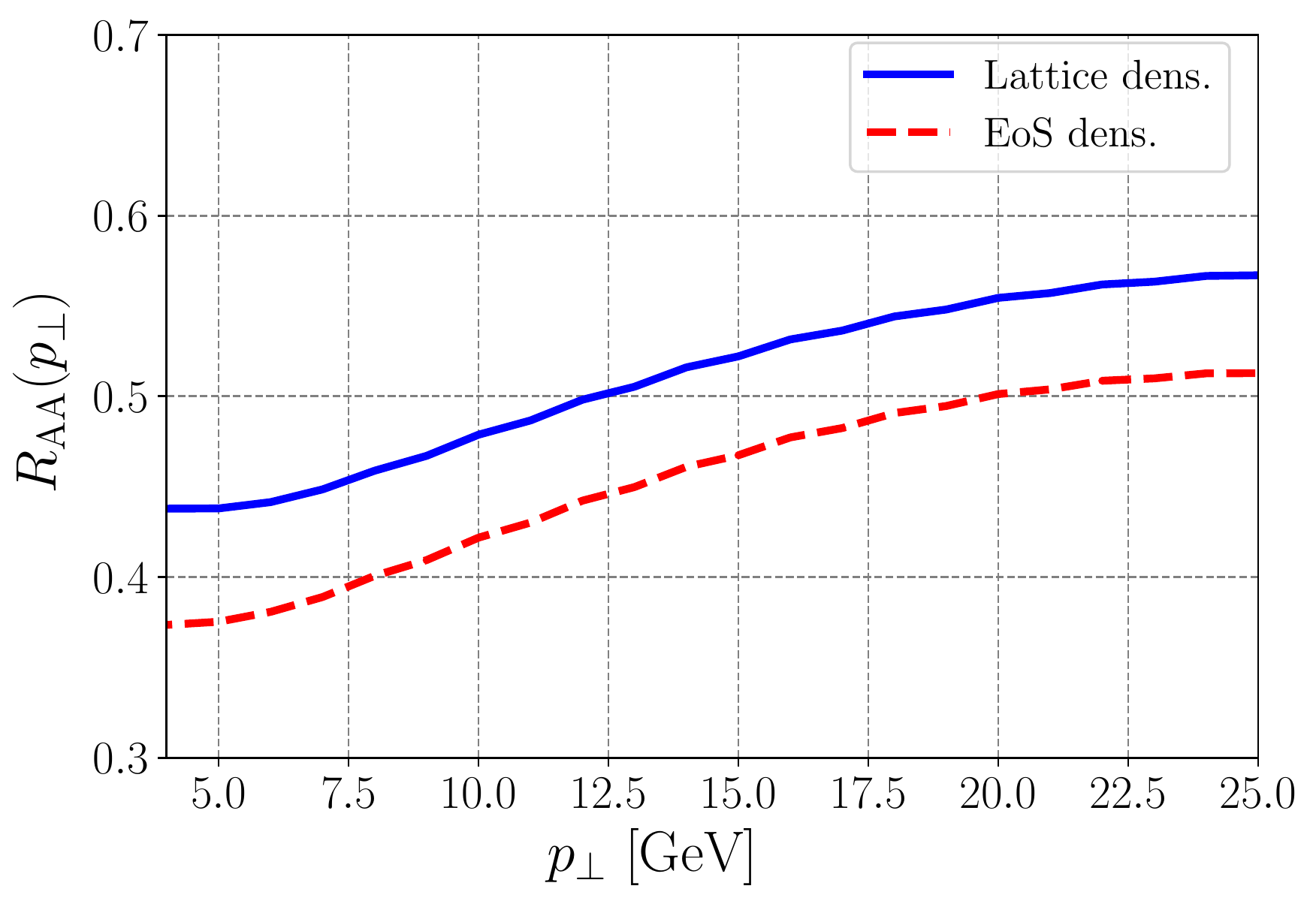}
\caption{ (Color online.) Nuclear modification factor of neutral pions for 62.4 GeV Au-Au collisions. The (blue) solid curve is the result for monopole density from the lattice, and the (red) dashed curve is the result for monopole density from thermodynamics.}
\label{fig:raa_bes}
\end{center}
\end{figure}

\begin{figure}[th!]
\begin{center}
\includegraphics[width=.5\textwidth,angle=0]{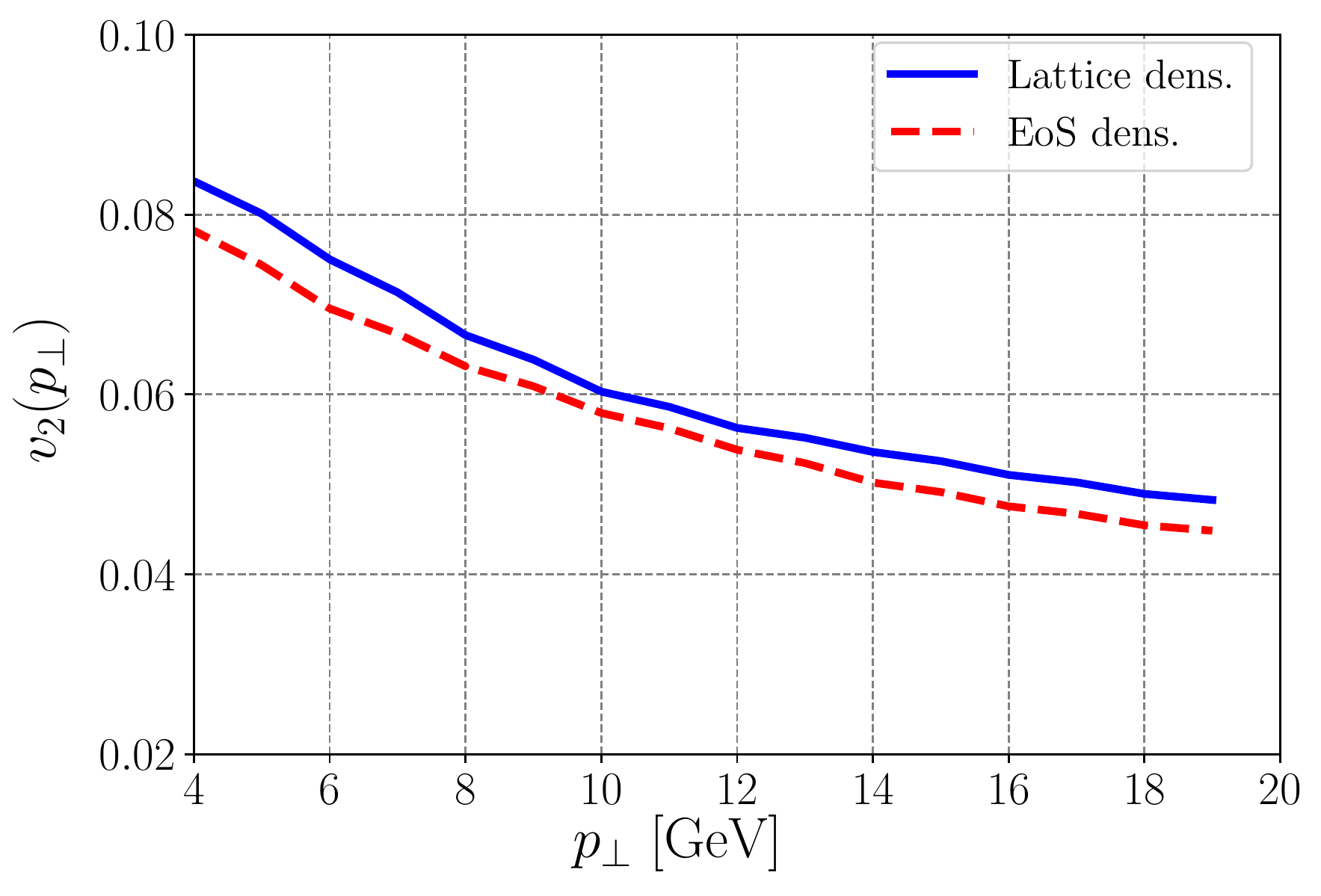}
\caption{ (Color online.) Azimuthal anisotropy of neutral pions for 62.4 GeV Au-Au collisions. The (blue) solid curve is the result for monopole density from the lattice, and the (red) dashed curve is the result for monopole density from thermodynamics.}
\label{fig:v2_bes}
\end{center}
\end{figure}

\begin{figure}[th!]
\begin{center}
\includegraphics[width=.5\textwidth,angle=0]{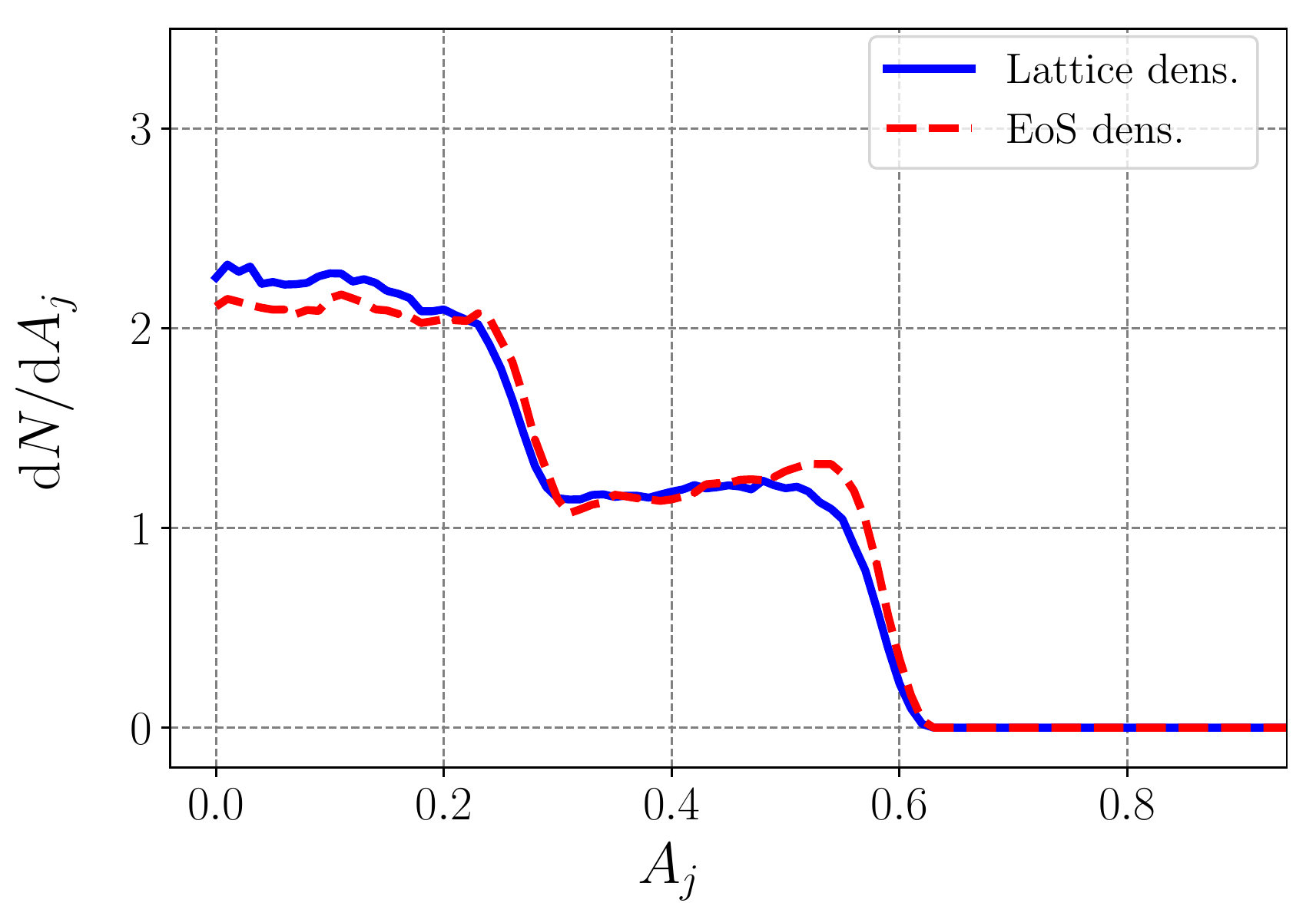}
\caption{ (Color online.) Dijet asymmetry of parton jets for 62.4 GeV Au-Au collisions. The (blue) solid curve is the result for monopole density from the lattice, and the (red) dashed curve is the result for monopole density from thermodynamics.}
\label{fig:dijet_bes}
\end{center}
\end{figure}

With the sPHENIX detector, RHIC will be able to detect jets in lower energy collisions. As stated earlier, monopole effects are strongest near $T_c$, and if we are able to study jet quenching in this lower energy range where most of the matter has a temperature of approximately $T_c$, we will get a better understanding of monopole features, such as the location and shape of the density peak.

In preparation for these experiments, we seek to make predictions -- with our jet quenching framework and Glauber initial conditions for a 62.4 GeV Au-Au collision -- for how jet observables will be altered when probing lower energy collisions. For the most realistic results, we would need a better understanding of the initial conditions and hydrodynamic expansion of the lower-energy fireball, and a better understanding of the initial energy spectrum of the produced parton jets.

Nevertheless, we can carry out the simulations to get relatively good predictions, the results of which are seen in Figs. \ref{fig:raa_bes} and \ref{fig:v2_bes} for $R_\text{AA}$ and  $v_2$, respectively. The results of jet quenching with monopoles at 62.4 GeV is very similar to that at 200 GeV; the $R_\text{AA}$ is slightly smaller and the $v_2$ is approximately the same. 

The deviation between the equation of state and directly-observed lattice monopole densities in the $R_\text{AA}$ is present in the 62.4 GeV Au-Au collision, as it was in the higher energy collisions -- particularly the 2.76 TeV Pb-Pb collision, where one of the densities did not agree with the data. On the other hand, the $v_2$ is not as sensitive to the monopole density scheme used, which was also the case for the higher energy collisions. The 62.4 GeV (and lower energy) runs at RHIC will help constrain the features of monopole density at temperatures near $T_c$, while the 2.76 TeV (and higher energy) collisions at the LHC are helpful in exploring the higher energy limit of the monopole density.

Without monopoles (not shown in the figures), $R_\text{AA}$ is unity and $v_2$ is approximately zero across all jet energies. This is due to the fact that at the temperatures produced in 62.4 GeV collisions -- the initial matter is slightly above $T_c$ --, the quark and gluon densities are small. If the number of quark and gluon degrees of freedom are proportional to the Polyakov loop, then, without monopoles, one would  expect very little nuclear modification of the parton jets in 62.4 GeV (and lower energy) collisions. 

However, the temperatures produced in these collisions are in the range where monopoles dominate. So, when including monopole contributions, we see a significant nuclear modification, comparable in magnitude to higher energy collisions. Therefore, if the data from the BES program show that there is significant medium-induced modification of parton jets, this will indicate that the near-$T_c$ scatterings on matter are strong. This data will also help constrain the shape of the monopole density curve and its peak, which, as we have shown, has a significant influence on the results of our model.

Fig. \ref{fig:dijet_bes} shows a predicted dijet asymmetry (with no cut) for the 62.4 GeV Au-Au collisions. As shown above, the Glauber initial conditions do not reproduce the LHC dijet asymmetry data well, while the fluctuating initial conditions have a decent agreement. Therefore, to accurately predict the 62.4 GeV Au-Au dijet asymmetry, we would need a more realistic, fluctuating hydrodynamic background with a $p_\perp$ cut to reflect the experimental cuts. Nevertheless, from comparison of Fig. \ref{fig:dijet_bes} with the results for the Glauber initial conditions at higher energies, we predict that the asymmetry peak in 62.4 GeV Au-Au collisions will not be as wide as it is in 200 GeV and 2.76 TeV collisions.

\section{Summary}
In this paper, we have described various aspects of jet quenching phenomena using the BDMPS formalism, including not only scattering on electric quasiparticles, quarks and gluons, but scattering on monopoles as well.  Unlike previous works by others, we include densities of all quasiparticles from certain common fit to lattice thermodynamics, so there are no free parameters in the theory.

The calculated observables include the nuclear modification factor $R_\text{AA}$, the azimuthal asymmetry $v_2$, and dijet asymmetry $A_j$; this is done both for RHIC and LHC energies. The main conclusion of the work is that the current model provides rather reasonable description of all of them. We find that, while realistic hydrodynamics is necessary for good agreement with the observables seen in experiment, account for event-by-event fluctuations is not necessary for reproducing $R_\text{AA}$ and $v_2$ data; fluctuations, however, seem to play a role in $A_j$. In all cases, the contribution of the monopoles is crucial for the success of the model.

In our model, monopoles give the dominant contributions to these observables at lower RHIC energies, e.g. 62.4 GeV, where quarks and gluons provide almost no quenching. Experimental observation of $R_\text{AA}$, $v_2$, and $A_j$ at lower energies that deviate from the $pp$ results would bolster the proposition of the magnetic scenario above $T_c$.

Still, the model in its current form has certain limitations, which need to be addressed in further studies. The model itself is that
of independent scattering on ``scatterers", which is of course an approximation. While we tried to remedy this partly by including monopole correlation corrections, clearly the system is a strongly coupled plasma, and more work 
is needed to include scattering effects more accurately. 

 We realize that the model we are using is missing some physical processes, as it only has radiative effects to lowest order, and neglects elastic and quasi-elastic scattering which believed to be necessary for quenching of jets with heavy $c,b$ quarks. Note that the recoil energy by scatterers is neglected; this, however, is only true if scatterers are heavy.
In fact, the efficiency of such processes depend on quasiparticle masses, and that in the near-$T_c$ region the monopoles are believed to be the lightest ones, thus contributing more to the elastic energy loss.  We hope to return to the issue and include some of those effects in subsequent works.

 \vskip 1cm

{\bf Acknowledgements.}
The authors thank Jean-Fran\c{c}ois Paquet for useful discussions and for providing a sample of the hydrodynamic backgrounds generated by the MUSIC code that we have used for our study. The authors also thank the Institute for Advanced Computational Science (IACS) at Stony Brook University for the time at its  LI-red computational cluster.
This work was supported in part by the U.S. D.O.E. Office of Science,  under Contract No. DE-FG-88ER40388.

 \appendix 
 
 \begin{figure*}[th!]
\begin{center}
\includegraphics[width=\textwidth,angle=0]{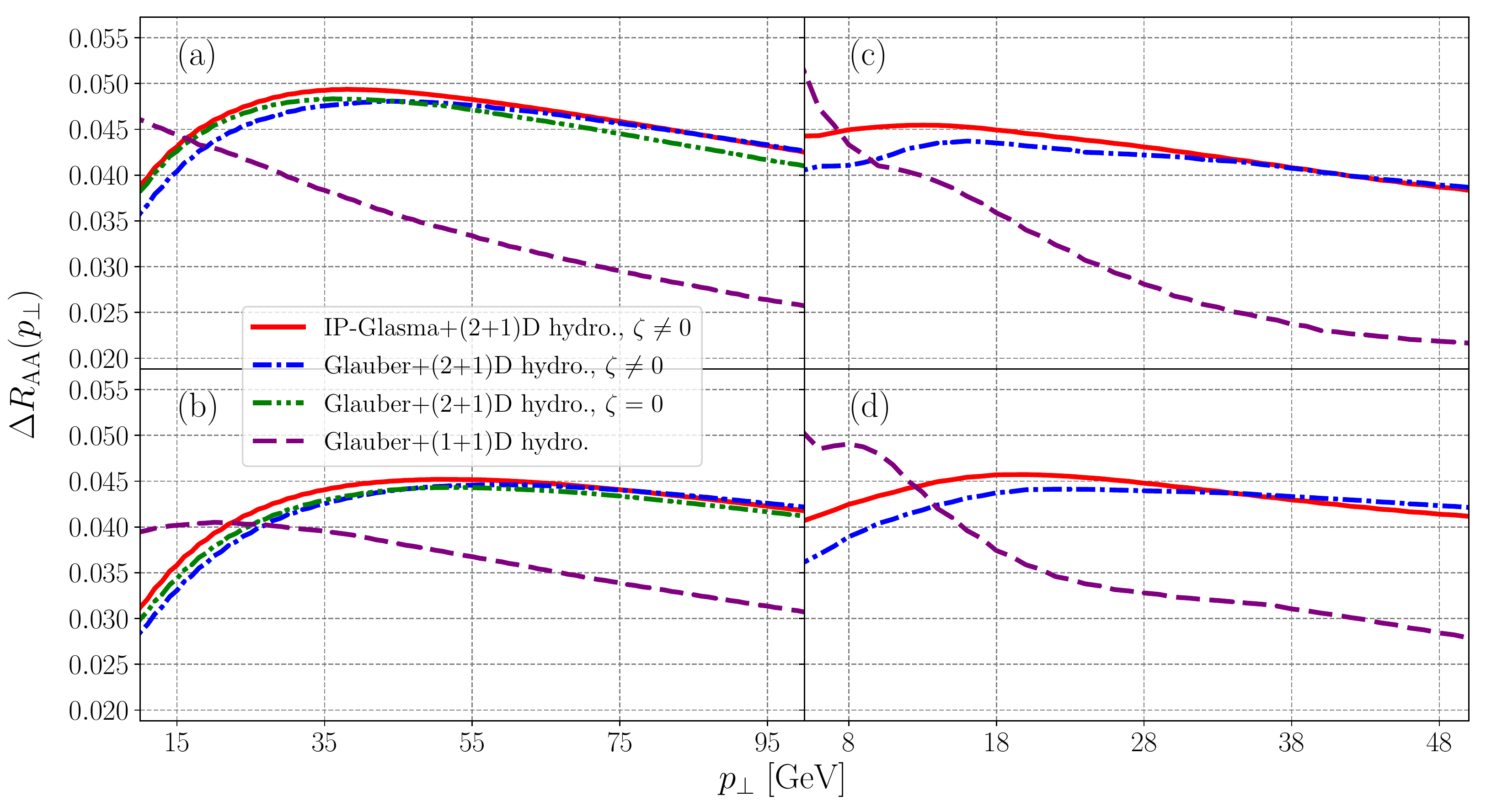}
\caption{(Color online.) The effect of monopole correlations on the nuclear modification factor of charged hadrons in 2.76 Pb-Pb collisions (a),(b), and neutral pions in 200 GeV Au-Au collisions (c),(d). The first row (a),(c) is the results for monopole density from the lattice, while the second row (b),(d) is results for monopole density from the equation of state. The (red) solid curve is for IP-Glasma initial conditions and (2+1)D hydrodynamics with bulk viscosity ($\zeta \neq 0$), the (blue) dash-dot curve is for Glauber initial conditions and (2+1)D hydrodynamics with bulk viscosity ($\zeta \neq 0$), the (green) dash-dot-dot curve is for Glauber initial conditions and (2+1)D hydrodynamics without bulk viscosity ($\zeta = 0$), and the (purple) dashed curve is for the smooth Glauber initial condition with  (1+1)D Bjorken evolution.}
\label{fig:raa_corr}
\end{center}
\end{figure*}

\begin{figure*}[th!]
\begin{center}
\includegraphics[width=\textwidth,angle=0]{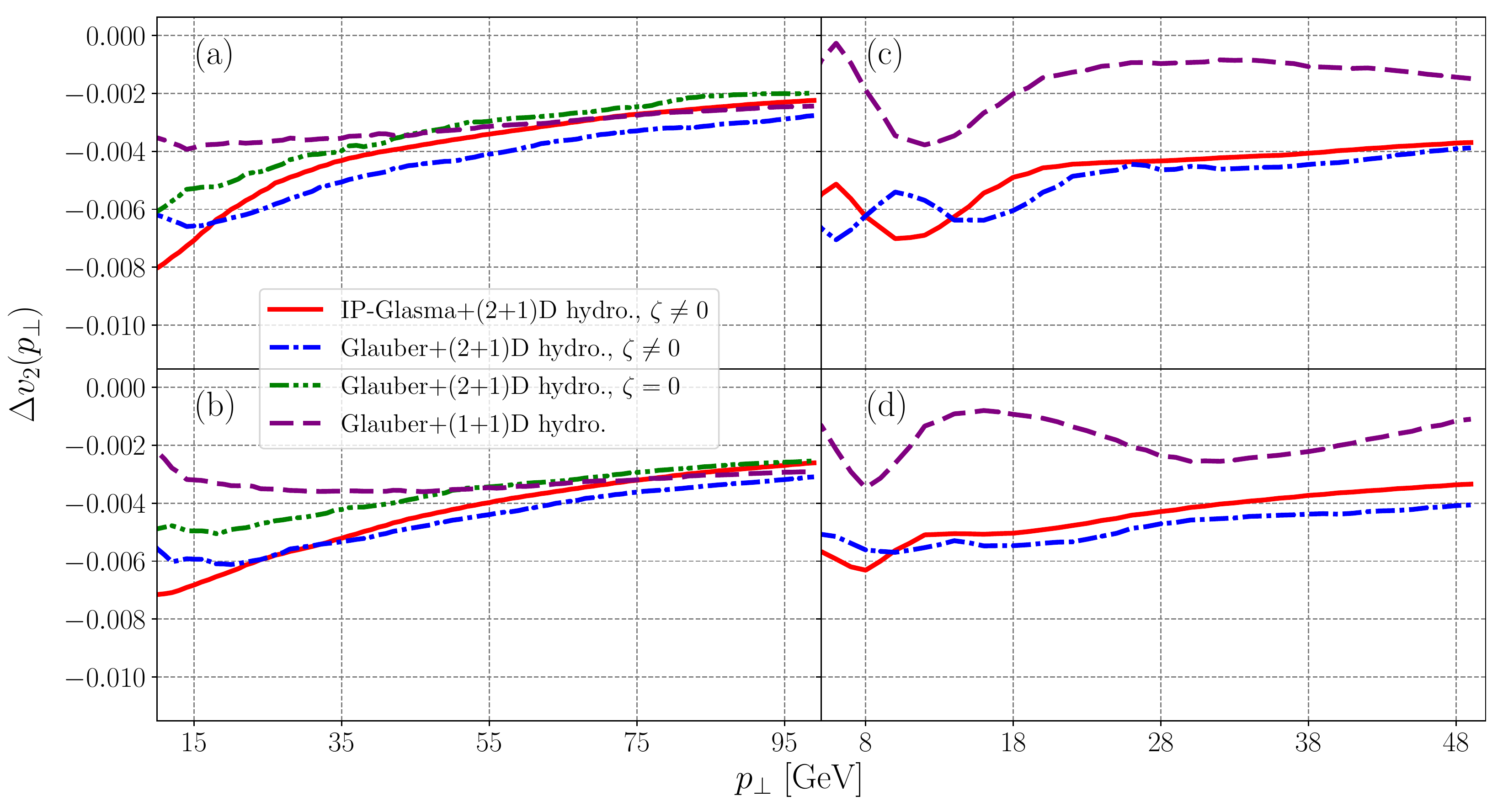}
\caption{(Color online.) The effect of monopole correlations on the azimuthal anisotropy of charged hadrons in 2.76 Pb-Pb collisions (a),(b), and neutral pions in 200 GeV Au-Au collisions (c),(d). The first row (a),(c) is the results for monopole density from the lattice, while the second row (b),(d) is results for monopole density from the equation of state. The (red) solid curve is for IP-Glasma initial conditions and (2+1)D hydrodynamics with bulk viscosity ($\zeta \neq 0$), the (blue) dash-dot curve is for Glauber initial conditions and (2+1)D hydrodynamics with bulk viscosity ($\zeta \neq 0$), the (green) dash-dot-dot curve is for Glauber initial conditions and (2+1)D hydrodynamics without bulk viscosity ($\zeta = 0$), and the (purple) dashed curve is for the smooth Glauber initial condition with  (1+1)D Bjorken evolution.}
\label{fig:v2_corr}
\end{center}
\end{figure*}
 \section{Parametrizations} \label{app_param}
 
 In this Appendix, we present the parameterizations for the equation of state and quasiparticle densities used in our study.
 
  \subsection{Equation of state}
  
 The parametrization of the equation of state (pressure) was given in Ref. \cite{Bazavov:2014pvz} as,
\al{
p/T^4 = & \left(\frac{\frac{0.3419}{T^4}+\frac{3.92}{T^2}-\frac{8.7704}{T}+\frac{19 \pi ^2}{36}}{2
   \left(-\frac{0.0475}{T^4}+\frac{0.8425}{T^2}-\frac{1.26}{T}+1\right)} \right) \nn \\
   & \times  (\tanh (3.8706 (T-0.9761))+1)\,,
}
and the equation of state for the energy density from Ref. \cite{Bazavov:2014pvz} was fitted in Mathematica to be
\al{
\epsilon/T^4 = &\left(\frac{20.89\, +\frac{23.55}{T^4}-\frac{57.62}{T^3}+\frac{59.79}{T^2}-\frac{40.37}{T}}{2 \left(\frac{0.30}{T^4}+\frac{2.17}{T^3}-\frac{3.56}{T^2}+\frac{0.57}{T}+1\right)} \right)\nn \\
&\times (\tanh (1.17
   (T-1.25))+1)\,.
   \label{eq_endens}
}

 \subsection{Densities of quarks, gluons, and monopoles near $T_c$}\label{app_rho}
 The density of magnetic monopoles is taken from Ref. \cite{Bonati:2013bga}; the density of each type of monopole (of which there are two in $SU(3)$) is given by 
\eq{
\frac{\rho_m}{T^3} = \frac{3.66}{\log((1/0.163)T/T_c)^3}\,.
\label{eq_rho}
}
Ref. \cite{Liao:2008vj} suggests that the monopole density falls off quickly below the critical temperature. 

The densities of electric particles are found using the Fermi-Dirac distribution and equation of state (we choose to use the pressure, but one can, in principle, use the entropy instead), following Ref. \cite{Xu:2014tda}. The parametrization of the Polyakov loop is
\eq{
L(T) = \left(\frac{1}{2} \tanh (7.69 (0.155 T-0.0726))+\frac{1}{2}\right)^{10}\,,
} and 
Using the ansatz that $\rho_E(T) \propto  c_q L(T) + c_g L^2(T)$, we have that the densities of quarks and gluons, respectively, are given by,
\al{
\rho_q / T^3 =&\frac{1.71E{-4} \left((T (9.87 T-16.6)+7.43) T^2+0.648\right)}{T^2
   ((T-1.26) T+0.843)-0.0475}\nn \\ &\times (\tanh (3.87 (T-0.976))+1) \\&\times (\tanh (7.69 (T-0.0726))+1)^{10} \nn\,,
}
\al{
\rho_g / T^3 =& \frac{\left(8.48E{-8}\right) \left((T (9.87 T-16.6)+7.43) T^2+0.648\right)}{T^2 ((T-1.26) T+0.843)-0.0475} \nn \\ &\times (\tanh (3.87 (T-0.976))+1) \\ &\times (\tanh (7.69(T-0.0726))+1)^{20} \nn\,.
   }

\section{Hadronic observables in our hydrodynamic calculations}
\label{app_hydro_obs}

The hadronic observables produced in our hydrodynamic simulations with optical Glauber initial conditions and the experimental data are shown in Tables \ref{hydro624}, \ref{hydro200}, and \ref{hydro2760}, for 62.4 GeV Au-Au, 200 GeV Au-Au, and 2.76 TeV Pb-Pb collisions, respectively, for hydrodynamics with bulk viscosity; Table \ref{hydro2760b} shows the calculation for hydrodynamics without bulk viscosity \cite{Paquet:2017pc}. In general, we have good agreement with data. The $v_2$ of our calculations is smaller than that of experiment, which is a well-known result of using optical Glauber initial conditions.

\begin{table}[h!]
\centering
\caption{Comparison of calculated and experimental hadronic observables in 62.4 GeV Au-Au collisions, for hydrodynamics with bulk viscosity. Data from Ref. \cite{Abelev:2008ab}.}
\label{hydro624}
\begin{tabular}{|l|l|l|}
\hline
                                & hydro. calc.       & data         \\
                                \hline
$N_\text{pion}$                           & 97.75     & 98.9$\pm$6.9    \\

$\braket{p_T}_\text{pion}$& 0.412     & 0.403$\pm$0.013 \\
$v_2(p_T=0.2-1.0)$                     & 0.054 & 0.0613       \\
$v_2(p_T=1.0-2.0)$                    & 0.163 & 0.132      \\ \hline
\end{tabular}
\end{table}

\begin{table}[h!]
\centering
\caption{Comparison of calculated and experimental hadronic observables in 200 GeV Au-Au collisions, for hydrodynamics with bulk viscosity. Experimental data is the same as used in Refs. \cite{Ryu:2015vwa,Ryu:2017qzn}.}
\label{hydro200}
\begin{tabular}{|l|l|l|}
\hline
                & hydro. calc.       & data         \\
                                \hline
$N_\text{pion}$                          & 133.4     & 135$\pm$10      \\
$\braket{p_T}_\text{pion}$& 0.422    & 0.411$\pm$0.021 \\
$v_2$                              & 0.059 & 0.0642 \\
\hline    
\end{tabular}
\end{table}

\begin{table}[h!]
\centering
\caption{Comparison of calculated and experimental hadronic observables in 2.76 TeV Pb-Pb collisions, for hydrodynamics with bulk viscosity. Experimental data is the same as used in Refs. \cite{Ryu:2015vwa,Ryu:2017qzn}.}
\label{hydro2760}
\begin{tabular}{|l|l|l|}
\hline
                & hydro. calc.       & data         \\
                                \hline
$N_\text{pion}$                           & 309.1     & 307$\pm$20        \\
$\braket{p_T}_\text{pion}$& 0.508    & 0.512$\pm$0.017   \\
$v_2$                            & 0.0746 & 0.0831$\pm$0.0034 \\ 
\hline
\end{tabular}
\end{table}

\begin{table}[h!]
\centering
\caption{Comparison of calculated and experimental hadronic observables in 2.76 TeV Pb-Pb collisions, for hydrodynamics without bulk viscosity. Experimental data is the same as used in Refs. \cite{Ryu:2015vwa,Ryu:2017qzn}.}
\label{hydro2760b}
\begin{tabular}{|l|l|l|}
\hline
                & hydro. calc.       & data         \\
                                \hline
$N_\text{pion}$                           & 299.079     & 307$\pm$20        \\
$\braket{p_T}_\text{pion}$& 0.616    & 0.512$\pm$0.017   \\
$v_2$                            & 0.075 & 0.0831$\pm$0.0034 \\ 
\hline
\end{tabular}
\end{table}

\section{Effects of monopole correlations on observables}
 \label{app_corr}

The preceding results were all computed using a monopole correlation factor of 0.85 to account for the change in $\hat{q}$ due to correlations of monopoles. In Figs. \ref{fig:raa_corr} and \ref{fig:v2_corr}, we show the effects of this correlation factor on $R_\text{AA}$ and $v_2$, respectively, for 2.76 TeV and 200 GeV collisions. We define, \eq{\Delta(\text{Obs.}) = (\text{Obs.})_\text{corr.} - (\text{Obs.})_\text{uncorr.}\,,} where (Obs.) is an observable.

We see that, in general, correlations of monopoles cause the $R_\text{AA}$ to increase and the $v_2$ to decrease. Also, we see that, for all realistic initial conditions and hydrodynamic evolutions, the monopole correlations have approximately the same effect, and that the magnitude of the effect is on the order of 10\%.

\end{document}